\documentclass[conference]{IEEEtran}
\IEEEoverridecommandlockouts
\usepackage{graphicx}
\usepackage{tabularx}
\usepackage[justification=centering]{caption}
\usepackage{cite}
\usepackage{url}
\usepackage{gensymb}
\usepackage{gensymb}
\usepackage{graphicx}
\usepackage{tabularx}
\usepackage[justification=centering]{caption}
\usepackage{cite}
\usepackage{latexsym}
\usepackage{amsmath}
\usepackage{algorithm}
\usepackage[noend]{algpseudocode}
\usepackage{booktabs}
\usepackage{multirow}
\usepackage{siunitx}
\usepackage{pdfpages}
\usepackage{array,booktabs}
\usepackage{amsmath}
\usepackage[utf8]{inputenc}
\usepackage{mathtools}
\usepackage{amssymb}
\usepackage{pdfpages}
\usepackage{booktabs}
\usepackage{afterpage}
\usepackage{float}
\usepackage{bbding}
\usepackage{pifont}
\usepackage{wasysym}
\usepackage{subfigure}
\usepackage{amsfonts}
\usepackage{siunitx}
\usepackage{xcolor}
\usepackage{mathtools, cases}
\usepackage{float}
\usepackage{stfloats}
\usepackage{caption}
\newcolumntype{P}[1]{>{\centering\arraybackslash}p{#1}}
\newcolumntype{M}[1]{>{\centering\arraybackslash}m{#1}}
\DeclareCaptionLabelFormat{continued}{#1~#2}
\renewcommand{\baselinestretch}{.97}
\ifCLASSOPTIONcompsoc
  \usepackage[nocompress]{cite}
\else
  \usepackage{cite}
\fi

\ifCLASSINFOpdf
\else
\fi

\begin{document}
%
\title{A Survey on Low Latency Towards 5G: RAN, Core Network and Caching Solutions }

\author{\IEEEauthorblockN{Imtiaz Parvez, \IEEEmembership{Student Member, IEEE}, Ali Rahmati, \IEEEmembership{Student Member, IEEE}, Ismail Guvenc, \IEEEmembership{Senior Member, IEEE},\\  Arif I. Sarwat, \IEEEmembership{Senior Member, IEEE}, and Huaiyu Dai, \IEEEmembership{Fellow, IEEE}}}
%
%

\maketitle

\begin{abstract}
	
The fifth generation (5G) wireless network  technology is to be standardized by 2020, where main goals are to improve capacity, reliability, and energy efficiency, while reducing latency and massively increasing connection density. An integral part of 5G is the capability to transmit touch perception type real-time communication empowered by applicable robotics and haptics equipment at the network edge. In this regard, we need drastic changes in network architecture including core and radio access network (RAN) for achieving end-to-end latency on the order of 1 ms. In this paper, we present a detailed survey on the emerging technologies  to achieve low latency communications considering three different solution domains: RAN, core network, and caching. We also present a general overview of 5G cellular networks composed of software defined network (SDN), network function virtualization (NFV), caching, and mobile edge computing (MEC) capable of meeting latency and other 5G requirements.


\end{abstract}


\section {Introduction}

The focus of next generation mobile communication is to provide seamless communication for machines and devices building the Internet-of-Things (IoT) along with personal communication. New applications such as tactile Internet\footnote{A network or network of networks for remotely accessing, perceiving, manipulating or controlling real or virtual objects or processes in perceived real time by humans or machines \cite{TactileInterent}.}, high-resolution video streaming, tele-medicine, tele-surgery, smart transportation, and real-time control dictate new specifications for throughput, reliability, end-to-end (E2E) latency, and  network robustness \cite{immerge}. Additionally, intermittent or always-on type connectivity is required for machine-type communication (MTC) serving diverse applications including sensing and monitoring, autonomous cars, smart homes, moving robots and manufacturing industries.

Several emerging technologies including  wearable devices, virtual/augmented reality, and full immersive experience (3D) are shaping the demeanor of human end users, and they have special requirements for user satisfaction. Therefore, these use cases of the next generation network push the specifications of 5G in multiple aspects such as data rate, latency, reliability, device/network energy efficiency, traffic volume density, mobility, and connection density. Current fourth generation (4G) networks are not capable of fulfilling all the technical requirements for these services.

Fifth generation (5G) cellular network is the wireless access solution to fulfill the wireless broadband communication specifications of 2020 and beyond \cite{SCMA2,Survey11}. In ITU, 5G ITU-R working group is working for development of 5G under the term IMT-2020 \cite{ITU-R}. The vision of this work is  to achieve one thousand times throughput improvement, 100 billion connections, and close to zero latency \cite{immerge,SCMA2}. In particular, 5G will support enhanced mobile broadband (MBB) with end-user data rates of 100~Mbps in the uniform spatial  distribution with peak data rates of 10-20 Gbps~\cite{Survey11,SCMA2}. Based on consensus, 5G will not only provide personal mobile service, but also massive machine type communications (MTC), and latency/reliability critical services. {\color{black}In mission critical communication (MCC)/ultra reliable low latency communication (uRLLC\footnote{uRLLC allows E2E latency \textcolor{black}{of} less than 1~ms on all layers with packet error rates of $10^{-5}$ to $10^{-9}$.}), both the \emph{latency} and reliability issues need to be addressed  \cite{publicsafety, ji2017introduction}.} In many cases, the corresponding E2E latency as low as 1 ms needs to be met with  reliability as high as $99.99\%$ \cite{7504504}. 


To achieve low \emph{latency} for MCC, drastic changes in the network architecture need to be performed. Since the delay is contributed by radio access network (RAN) and core network along with backhaul between RAN and core network, new network topology involving software define network (SDN), network virtualized function (NFV), and mobile edge computing (MEC)/caching can be employed to reduce the latency significantly. This can happen due to the capability of  seamless operation and independence from hardware functionality provided by these entities. Moreover, new physical air interface with small time interval transmission, small size packets, new waveforms, new modulation and coding schemes are the areas of investigation for attaining low latency. In addition, optimization of radio resource allocation, massive MIMO, carrier aggregation in millimeter wave, and priority of data transmission need to be addressed. All in all, a robust integration with existing LTE is necessary for  5G networks that will enable industries to deploy 5G quickly and efficiently when it is standardized and available. In summary, 5G wireless access should be an evolution of LTE complemented with revolutionary architecture designs and radio technologies.

\textcolor{black}{Even though the goals of 5G are ambitious based on 4G point of view, researchers from industry and academia are working to bring 5G key performance indicator (KPI) goals (including low latency) into reality. The 5G road map is fixed: 5G standardization is set up by 2018, 5G first commercial launch is to be by 2020 and 5G worldwide launch will be  by 2022 and onwards \cite{5Groadmap}. Along with the ITU, various European research projects such as METIS-II, MiWaveS and mmMAGIC are working  to address diverse aspects of 5G such as RF block and algorithm for mmWave communication. On the other hand, projects such as ADEL, FANTASTIC 5G, SPEED 5G, and Flex5GWare address hardware and fundamental building blocks for 5G, while 5G CHAMPION European/Korean research project is working  to implement  proof of concept (PoC) of 5G network encompassing all cutting edge radio, core and satellite technologies~\cite{7848798}. It aims to showcase the 5G PoC with \emph{latency} of 1 ms  on 2018 Winter Olympics in PyeongChang, Korea. Telecommunication vendors such as Ericsson, Huawei and Nokia Siemens are working to bring  network infrastructure and UE for 5G roll out by 2020. Besides, \textcolor{black}{researchers} from academia are working on different aspects/goals of 5G including low latency. \textcolor{black}{However, the real field PoC and benchmarking of performance is to be done.}}

In the literature, surveys on 5G network including architecture \cite{SCMA2,Survey11}, SDN/NFV/MEC based core network \cite{SDNFVsurvey, MECSurvey}, caching \cite{cachingsurvey1,cachingsurvey2}, backhaul \cite{backhaulsurbey}, resource management \cite{ResourcemanSurvey} and data centric network \cite{DCNSurvey1, DCNSurvey2} are available. Apart from that, surveys on latency reduction approaches in Internet \cite{InterSurvey}, cloud computing \cite{cloudsurvey, 7997242} and distributed network applications \cite{Networksurvey1,Networksurvey2} are also presented; however, to the best of our knowledge, a comprehensive survey on latency reduction approaches in cellular networks towards 5G is not available yet. 

In this paper, we present a comprehensive survey of latency reduction solutions particularly in the context of 5G wireless technology. We first present the sources  and fundamental constraints for achieving \emph{low latency}  in a cellular network. We also overview an exemplary 5G network architecture with compliance to \textcolor{black}{IMT-2020 vision}. Finally, we provide an extensive review of proposed solutions for achieving low latency towards 5G. The goal of our study is to bring all existing solutions on the same page along with future research directions. We divide the existing solutions into three parts: (1) RAN solutions;  (2) Core network solutions;  (3) Caching solutions. However, detailed comparison of these solutions are beyond the scope of this work.

\begin{table}[h!]
	\centering
	    \footnotesize
		\caption{\textsc{List of Acronyms.}}
		\label{acronyms}	
		\begin{tabular}{|l|p{6cm}|} 
			\hline  AS & Access Stratum\\
			\hline AR & Augmented Reality \\
			\hline BC & Broadcast Channel\\
			\hline BLER & Block Error Rate\\
			\hline CCP & Communication Control Port\\
			\hline CCSE & Control Channel Sparse Encoding\\ 
			\hline CSIT & Channel State Information at the Transmitter\\
			\hline DTB & Delivery Time per Bit \\
			\hline  D2D & Device to Device \\
			\hline DRB & Data Radio Bearer\\
			\hline eMBB & Enhanced Mobile Broadband\\ 
			\hline EPC & Evolved Packet Core\\
			\hline FFT & Fast Fourier Transform\\
			\hline FDT & Fractional Delivery Time\\
			\hline FBMC &  Filter Bank Multi Carrier \\ 
			\hline GFDM & Generalized Frequency Division Multiplexing\\
			\hline GP & Guard Period\\
			\hline GTP & GPRS Tunnel Protocol\\
			\hline GGSN & Gateway GPRS (General Packet Radio Service) Service Node \\
			\hline HARQ & Hybrid Automatic-Repeat-Request\\
			\hline ICI & Inter Carrier Interference \\
			\hline  IoT & Internet of Things \\ 
			\hline ITS & Intelligent Transportation System\\ 
			\hline IFFT & Inverse Fast Fourier Transform\\ 
			\hline  ITU & International Telecommunications Union\\ 
			\hline ITU-R & International Telecommunication Union-Radio Communication Sector \\
			\hline IMT-2020 &  International Mobile Telecommunication System with a target date set for 2020 \\ 
			\hline ISI & Inter Symbol Interference \\ 
			\hline  MEC & Mobile Edge Computing\\ 
			\hline  MTP & Machine Type Communication \\ 
			\hline METIS & 	Mobile and Wireless Communications Enablers for Twenty-Twenty (2020) Information Society\\ 
			\hline NGMN & 	Next Generation Mobile Networks\\ 
			\hline MCC & Mission Critical Communication \\
			\hline MRC-ZF & Maximum Ratio Combining Zero Forcing\\ 
			\hline MAC & Medium Access Control\\
			\hline mMTc& Massive Machine Type Communication\\
			\hline MME & Mobility Management Entity\\
			\hline MIMO & Multiple Input Multiple Output\\
			\hline MDP & Markov Decision Process\\
			\hline  NFV & Network Function Virtualization\\ 
			\hline NAS & Non-Access Stratum\\
			\hline NDT & Normalized Delivery Time\\
			\hline  OFDM & Orthogonal Frequency Division Multiplexing\\
			\hline OOB & Out of Band\\ 
			\hline OW & Optical Window\\
			\hline OLLA & Outer Loop Link Adaptation \\
			\hline PUCCH& Physical Uplink Control Channel\\
			\hline P2P & Peer-to-Peer\\
			\hline RAN & Radio Access Network\\ 
			\hline SC-FDMA & Single Carrier Frequency Division Multiple Access\\
			\hline SRB & Signaling Radio Bearer\\
			\hline SCMA & Sparse Code Multiple Access\\
			\hline SGW & Serving GPRS Gateway\\
			\hline TDD & Time Division Duplex \\
			\hline RAT & Radio Access Technology \\ 
			\hline uRLLC & ultra Reliable Low Latency Communication\\
			\hline UFMC & Universal Filtered Multi-Carrier \\ 
			\hline UDN & Ultra Dense Network\\	 
			\hline VR & Virtual Reality \\ 
			\hline VLC & Visible Light Communication\\
			\hline ZF & Zero Forcing\\ 
			\hline 5G & Fifth Generation Mobile Network \\ 
			\hline 5GETLA & 5G Flexible TDD based Local Area \\ \hline 	
		\end{tabular} 
\end{table}
The rest of the paper is organized as follows. \textcolor{black}{Section~II presents the latency critical services in 5G. \textcolor{black}{The sources of latency in a cellular network are discussed in Section~III. Section~IV reviews the fundamental constraints and approaches for achieving low latency}. Three key low latency solutions in RAN, core network, and caching are presented in Sections V, VI, and VII, respectively.  Section~VIII presents the field tests, trials and experiments of low latency approaches. Open issues, challenges and future research directions are discussed in Section~IX.} Finally, concluding remarks are provided in Section~X. \textcolor{black}{Some of the acronyms  used in this paper are presented in Table~\ref{acronyms}.}

\section{Low Latency Services in 5G}
Latency is highly critical in some applications such as automated industrial production, control/robotics, transportation, health-care, entertainment, virtual realty, education, and culture. \textcolor{black}{In particular, IoT is quickly becoming a reality which connects anything to any other thing anytime, and anywhere. Smart wearable devices (smart watches, glasses, bracelets, and fit bit), smart home appliances (smart meters, fridges, televisions, thermostat), sensors, autonomous cars, cognitive mobile devices (drones, robots, etc.) are connected to always-on hyper-connected world to enhance our life style \cite{schulz2017latency,palattella2016internet,Sarwat2018}. Even though operators are supporting these IoT applications through existing 3G/LTE, some applications require much more stringent requirements from underlying networks such as low latency, high reliability {\cite{tavana2018congestion,7132976}}, and security \cite{aidin,7462517}.} In some cases, we need \emph{latency}  as low as 1 ms with packet loss rate no larger than $10^{-2}$. Several latency critical services which need to be supported by 5G  are described as follows.

\begin{itemize}
	\item {\textbf {Factory Automation:}} Factory automation includes real-time control of machine and system for quick production lines and limited human involvement. In these cases, the production lines might be numerous and contiguous. This is highly challenging in terms of latency and reliability. Therefore, the \textcolor{black}{E2E} \emph{latency}  requirement for factory automation applications is between 0.25~ms to 10~ms with a packet loss rate of $10^{-9}$\cite{Transportation2,factoryautamion2}. \textcolor{black}{In factory
		automation, the latency is measured as E2E in which
		the sensors measuring data are at one end and transmit
		the data for processing to the other end for programmable logic controller
		(PLC). The proposed values for the latency are based on the KoI (Koordinierte Industriekommunikation) project, in which
		a detailed questionnaire-based survey
		is conducted to collect the information from an extensive  range of
		factory automation processes \cite{holfeld2016wireless}.}
	
	\item {\textbf {Intelligent Transportation Systems:}} Autonomous driving and optimization of road traffic requires ultra reliable low latency communication. According to intelligent transportation systems (ITS), different cases including autonomous driving, road safety, and traffic efficiency  services have different requirements \cite{Transportation1,Transportation2}. Autonomous vehicles require coordination among themselves for actions such as platooning and overtaking \cite{plattoning}. \textcolor{black}{ For automated vehicle overtaking, maximum E2E latency of 10~ms is allowed for each message exchange. For video integrated  applications such as \emph{see-through-vehicle} application described in~\cite{7117947} requires to transmit raw video which allows maximum delay of 50~ms~\cite{7912382}.} Road safety includes warnings about  collisions or dangerous situations. Traffic efficiency services control traffic flow using the information of the status of traffic lights and local traffic situations. For these purposes, \emph{latency}  of 10~ms to 100~ms with packet loss rate of $10^{-3}$ to $10^{-5}$ is required.
	
	\item {\textbf {Robotics and Telepresence:}} In the near future, remote controlled robots will have applications in diverse sectors such as construction and maintenance in dangerous areas. A prerequisite for the utilization of robots and telepresence applications is remote-control with real-time synchronous visual-haptic feedback. In this case, system response times should be less than a few milliseconds including network delays \cite{instance1290, Transportation2,7564647}. Communication infrastructure capable of proving this level of real-time capacity, high reliability/availability, and mobility support is to be addressed in 5G networks.
	
\begin{table*}[!tp]
			\centering
			\caption{\textsc{\textcolor{black}{Typical Latency and Data Rate Requirements for Different Mission Critical Services.}}}
			\label{mylabellatencycri}	
			\begin{tabular}{p{3 cm}|p{02.5cm}|p{02.8cm}|p{7 cm}}
				\hline \textbf{Use case} & \textbf {\textcolor{black}{Latency}} &\textbf{\textcolor{black}{Data rate}}& \textbf{\textcolor{black}{Remarks}}\\ \hline
				\hline Factory Automation & 0.25-10~ms~\cite{Transportation2}& 1~Mbps ~\cite{7247338} & \begin{itemize}
				\vspace{-1.5mm}\item Generally factory automation applications require small data rates for motion and remote control.
				\item Applications such as machine tools operation may allow latency as low as 0.25~ms. \vspace{-2.5mm} \end{itemize}  \\ 
				\hline Intelligent Transport Systems (ITS)  & 10-100~ms~\cite{Transportation2} &
				10-700~Mbps ~\cite{Choi}& \begin{itemize}
				\vspace{-1.5mm} \item Road safety of ITS requires latency on the order of 10~ms. \item Applications such as virtual mirrors require data rates on the order of 700~Mbps. \vspace{-2.5mm} \end{itemize} \\ 
				\hline  Robotics and Telepresence &  1~ms~\cite{Telecpresence} &100~Mbps~\cite{NakiaWebnair}&  \begin{itemize}
				\vspace{-1.5mm}\item Touching an object by a palm may require latency down to 1 ms. \item VR haptic feedback requires data rates on the order of 100~Mbps.\vspace{-2.5mm} \end{itemize} \\ 
				\hline  Virtual Reality (VR) &  1~ms~\cite{instance1290} &1~Gbps~\cite{NakiaWebnair}& \begin{itemize}
				\vspace{-1.5mm}\item Hi-resolution $360^{\degree}$ VR requires high rates on the order of 1~Gbps while allowing latency of 1~ms.\vspace{-2.5mm} \end{itemize}  \\ 
				\hline  Health care & 1-10~ms~\cite{7564647} &100~Mbps~\cite{NakiaWebnair}& \begin{itemize} \vspace{-1.5mm} \item Tele-diagnosis, tele-surgery and tele-rehabilitation may require latency on the order of 1~ms with data rate of 100~Mbps.\vspace{-2.5mm} \end{itemize}\\ 
				\hline  Serious Gaming &  1~ms~\cite{instance1290} &1~Gbps~\cite{NakiaWebnair}& \begin{itemize}
				\item \vspace{-1.5mm} Immersive entertainment and human’s interaction with the high-quality visualization may require latency of 1~ms and data rates of 1~Gbps for high performance.\vspace{-2.5mm} \end{itemize} \\ 
				\hline  Smart Grid & 1-20~ms~\cite{instance1290, Transportation2} &10-1500~Kbps~\cite{smartgriddata}& \begin{itemize}
				\vspace{-1.5mm} \item Dynamic activation and deactivation in smart grid requires latency on the order of 1~ms. \item Cases such as wide area situational awareness require date rates on the order of 1500~Kbps.\vspace{-2.5mm} \end{itemize}\\ 
				\hline Education and Culture & 5-10~ms~\cite{instance1290} &1~Gbps~\cite{NakiaWebnair}& \begin{itemize}
				\vspace{-1.5mm}\item  Tactile Internet enabled multi modal human-machine interface may require latency as low as 5~ms. \item Hi-resolution $360^{\degree}$ and haptic VR may require data rates as high as 1~Gbps. \vspace{-2.5mm} \end{itemize}\\ 
				\hline 
			\end{tabular} 
		\end{table*}

	\item {\textbf {Virtual Reality (VR):}} Several applications such as micro-assembly and tele-surgery require very high levels of sensitivity and precision for object manipulations. VR technology accommodates such services where several users interact via physically coupled VR simulations in a shared haptic environment. Current networked communication does not allow sufficient low latency for stable, seamless coordination of users. Typical update rates of display for haptic information and physical simulation are in the order of 1000 Hz which allows round trip \emph{latency}  of 1~ms. Consistent local view of VR can be maintained for all users  if and only if the latency of around 1 ms is achieved \cite{7564647, instance1290,kasgari2018human}.

	\item {\textbf {Augmented Reality (AR):}} In AR technology, the augmentation of information into the user's field of view enables applications such as driver-assistance systems, improved maintenance, city/museum guides, telemedicine, remote education, and assistive  technologies for police and firefighters \cite{instance1290}. However, insufficient computational capability  of mobile devices and latency of current cellular network  hinder the applications. In this case, \emph{latency}  as low as a few milliseconds is required.

	\item {\textbf {Health care:}} Tele-diagnosis, tele-surgery and tele-rehabilitation are a few notable healthcare applications of low latency tactile Internet. These allow for remote physical examination even by palpation, remote surgery by robots, and checking of patients' status remotely. For these purposes, sophisticated control approaches with round trip latency of 1-10~ms and high reliability data transmission is mandatory \cite{7564647, instance1290}.

	\item {\textbf {Serious Gaming:}} The purpose of serious gaming is not limited to entertainment. Such games include problem-solving challenges, and goal-oriented  motivation which can have applications in different areas such as education, training, simulation, and health. Network \emph{latency}  of more than 30-50~ms results in a significant degrade in game-quality and game experience ratings. Ideally, a round trip time (RTT) on the order of 1~ms is recommended for perceivable human's interaction with the high-quality visualization \cite{instance1290}.
	
	\item {\textbf {Smart Grid:}} The smart grid has strict requirements of reliability and  latency~\cite{7011570,6939551,en9090691,8107405}. The dynamic control allows only 100~ms of E2E \emph{latency}  for switching suppliers (PV, windmill, etc.) on or off. However, in case of a synchronous co-phasing of power suppliers (i.e. generators), an E2E delay of not more than 1~ms is needed \cite{instance1290,SCMA2}. Latency more that 1~ms which is equivalent to a phase shift of about $18 \degree$ (50 Hertz AC network) or $21.6\degree$ (60 Hertz AC network), may have serious consequence in smart grid and devices.

	\item {\textbf {Education and Culture:}} Low latency tactile Internet will facilitate remote learning/education by haptic overlay of teacher and students. For these identical multi-modal human-machine interfaces, round trip latency of 5-10~ms is allowed for perceivable visual, auditory, and haptic interaction \cite{instance1290,7564647}. Besides that, tactile Internet will allow to play musical instruments from remote locations. In such scenarios, supporting network latency lower than few milliseconds becomes crucial \cite{instance1290}.
	
\end{itemize}

Based on the applications and use case scenarios above, latency critical services in 5G networks demand an E2E delay of 1~ms to 100~ms. \textcolor{black}{The latency requirements along with estimated data rates for various 5G services are summarized in Table~II. Some use cases such as VR and online gaming may require round trip latency on the order of 1~ms with data rates as high as 1~Gbps. On the other hand, use cases such as factory automation and smart grid require lower data rates on order of 1~Mbps with demanding latency of 1~ms. For required data rates on the order of 1~Gbps, \cite{NakiaWebnair} reports  that bandwidth of 40~MHz is sufficient at 20 node density per square kilometer. For data rates of few Mbps, bandwidth of 20~MHz and lower can be sufficient for most scenarios.} This means spectral efficiency supported by 5G is 50~bps/Hz while LTE-A can support upto 30~bps/Hz~\cite{7551912}. For lower bandwidth, spectrum below 6~GHz can be utilized while for high bandwidth requirement, mmWave can be an attractive choice~\cite{NakiaWebnair}.

In the next section, the major sources of latency in a cellular network are discussed.

%

\section{Sources of Latency in a Cellular Network }

In the LTE system, the \emph{latency}  can be divided into two major parts: (1) user plane (U-plane) latency and (2) control plane (C-plane) latency. The U-plane latency is measured by one directional transmit time of a packet to become available in the IP layer between evolved UMTS terrestrial radio access network (E-UTRAN) edge/UE and  UE/E-UTRAN node \cite{Latecyanaysis}. On the other hand, C-plane latency can be defined as the transition time of a UE to switch from idle state to active state. At the idle state, an UE is not connected with radio resource control (RRC). After the RRC connection is being setup, the UE switches from idle state into connected state and then enters into active state after moving into dedicated mode. Since the application performance is dependent mainly on the U-plane latency, U-plane is the main focus of interest for low latency communication.


\begin{figure}[!bp]
	\centering 
	\includegraphics[width=1\linewidth]{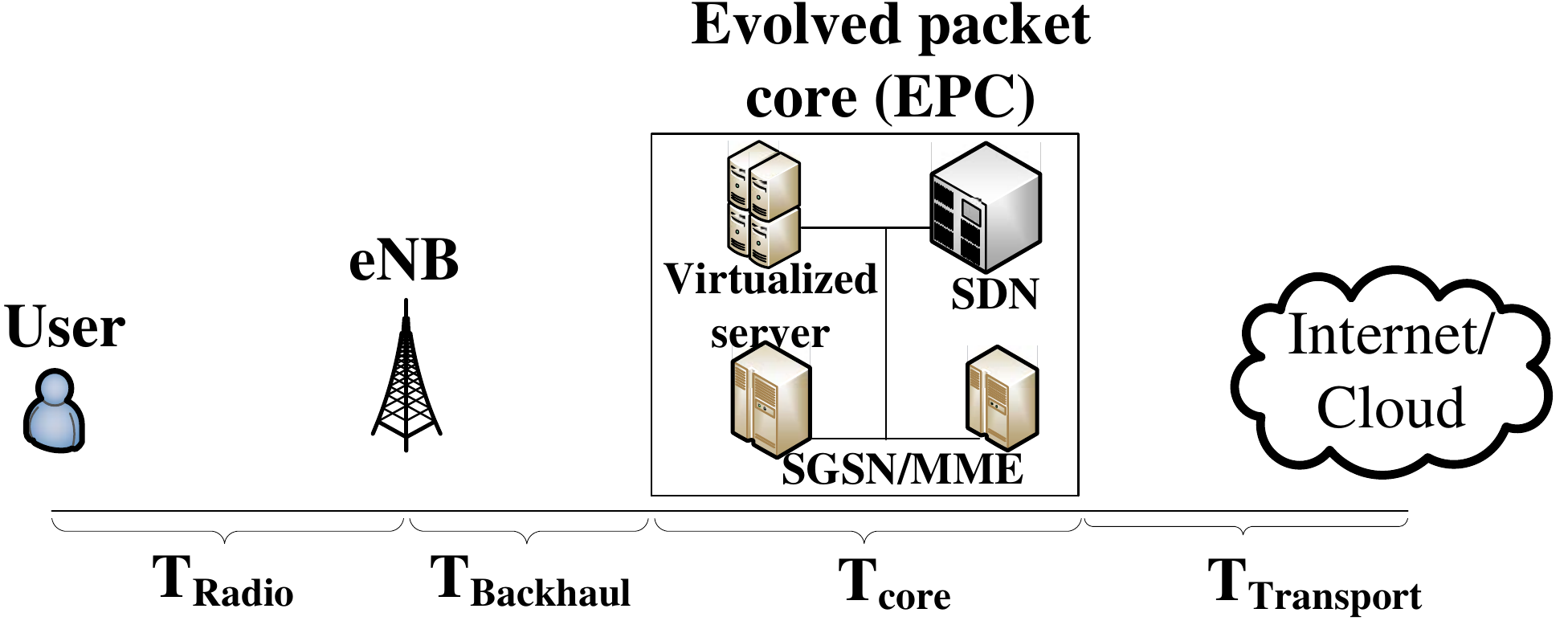}
	\caption{Latency contribution in E2E delay of a packet transmission.}
	\label{SourceLatency}
\end{figure}

\textcolor{black}{In the U-plane, the delay of a packet transmission in a cellular network can be contributed by the RAN, backhaul, core network, and data center/Internet.} As referred in Fig. \ref{SourceLatency}, the total one way transmission time \cite{EnablingLowLatency} of current LTE system can be written as 

\begin{equation}
T= T_{\text{Radio}} + T_{\text{Backhaul}} + T_{\text{Core}} + T_{\text{Transport}}
\end{equation}
where 
\begin{itemize}
\item $T_{\text{Radio}}$ is the  packet transmission time between eNB and UEs and is mainly due to physical layer communication. \textcolor{black}{It is contributed by eNBs, UEs and environment. It \textcolor{black}{consists of}  time to transmit, processing time at eNB/UE, \textcolor{black}{retransmissions, and}  propagation delay. Processing delay at the eNB involves channel coding, rate matching, scrambling, cyclic redundancy check (CRC) attachment, precoding, modulation mapper, layer mapper, resource element mapper\textcolor{black}{, and}  OFDM signal generation. On the other hand, uplink processing at UE involves CRC attachment, code block segmentation, code block concatenation, channel coding, rate matching,  data and control multiplexing, and channel interleaver. Propagation delay depends on obstacles (i.e. building, trees, hills etc.) on the way of propagation  and the total distance traveled by the RF signal;}

\item $T_{\text{Backhaul}}$ is the time for building connections between eNB and the core network (i.e. EPC). Generally, the core network and eNB  are connected by copper wires or microwave or optical fibers. \textcolor{black}{In general, microwave involves lower \textcolor{black}{latency}  while optic fibers come with comparatively higher latency. However, spectrum limitation may curb the capacity of microwave~\cite{wilner1976fiber};}

\item  $T_{\text{Core}} $ is the processing time taken by the core network. \textcolor{black}{It is contributed by various core network entities such as mobility management entity (MME), serving GPRS support node (SGSN), and SDN/NFV. The processing steps of core network includes NAS security, EPS bearer control, idle state mobility handling, mobility anchoring, UE IP address allocation, and packet filtering;}

\item  $T_{\text{Transport}}$ is the delay to data communication between the core network and Internet/cloud. \textcolor{black}{Generally, distance between the core network and \textcolor{black}{the server}, bandwidth, \textcolor{black}{and} communication protocol affect this latency;}

\end{itemize}
\begin{figure*}[!hbp]
	\centering
	\includegraphics[scale=0.85]{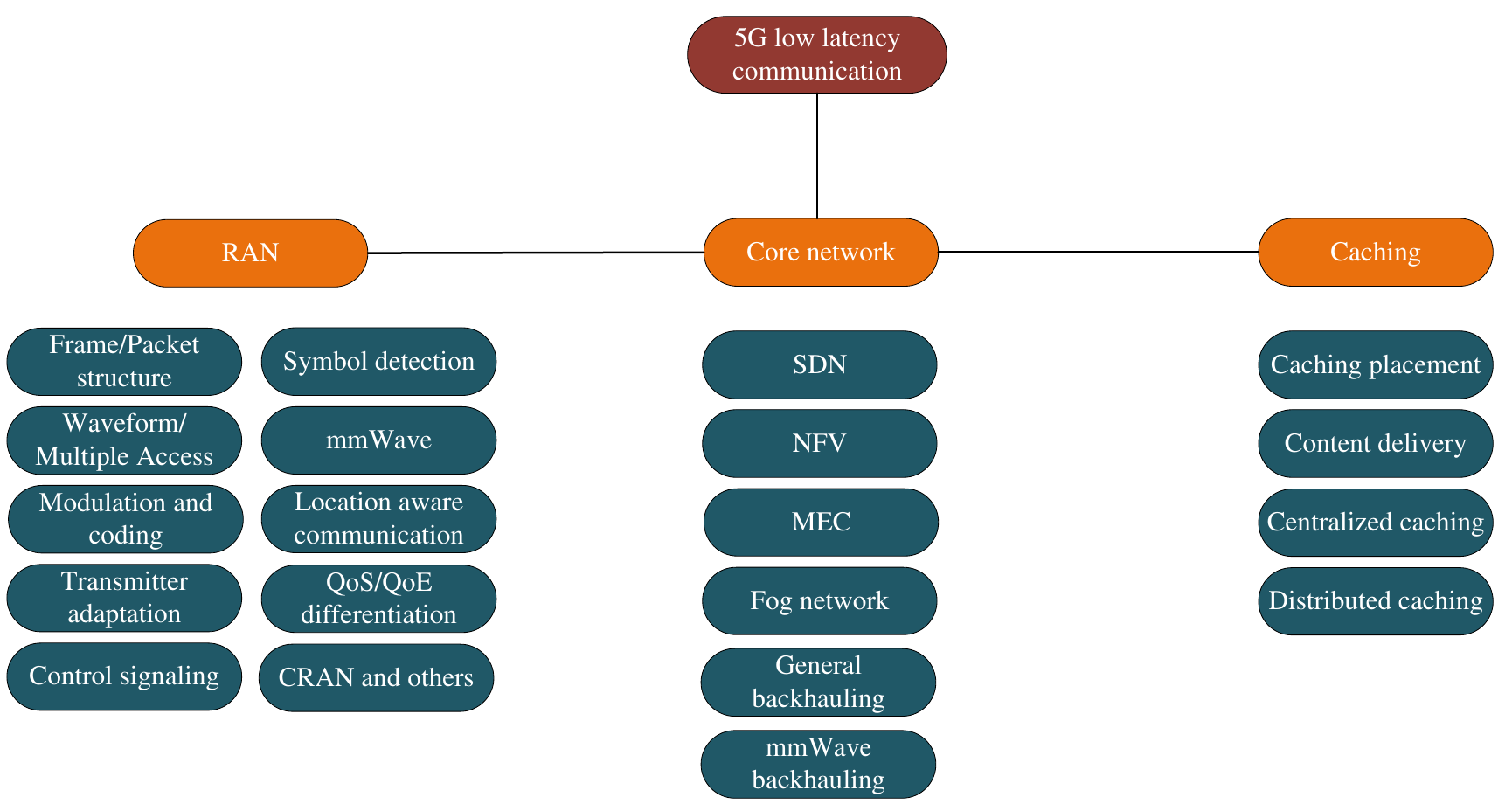}
	\caption{Categories of different solutions  for achieving low latency in 5G.}
	\label{CatagorySolutions}
\end{figure*}

The \textcolor{black}{E2E} delay, $T_{\mathrm{E2E}}$ is then approximately given by $2\times T$. The $T_{\text{Radio}}$ is the sum of   transmit time, propagation latency, processing time (channel estimation, encoding and decoding time for first time), and retransmission time (due to packet loss). In particular, the $T_{\text{Radio}}$ for a scheduled user \cite{pocovi2016impact,latenctrep} can be expressed as:

\begin{equation}
\textcolor{black}{T_\mathrm{Radio}=t_\mathrm{Q} + t_\mathrm{FA} + t_\mathrm{tx}+ t_\mathrm{bsp}+t_\mathrm{mpt}}
\end{equation}

where

\begin{itemize}
	\item  $t_\mathrm{Q}$ is the queuing delay which depends on the number of users that will be multiplexed on same resources;
	\item $t_\mathrm{FA}$ is the delay due to frame alignment which depends on the frame structure and duplexing modes (i.e., frequency division duplexing (FDD) and time division duplexing (TDD));
	\item $t_\mathrm{tx}$ is the time for transmission processing, and payload transmission which uses at least one TTI  depending on radio channel condition, payload size, available resources, transmission errors and retransmission;
	
\item  $ t_\mathrm{bsp}$ is the processing delay at the base station;
\item $t_\mathrm{mpt}$ is the processing delay of user terminal. Both the base station and user terminal delay depend on the capabilities of base station and  user terminal (i.e., UE), respectively.

\end{itemize}

In compliance with ITU, $T_{\text{Radio}}$ should not be more than 0.5~ms for \emph{low latency}  communication \cite{Designcretia}. In this regard, radio transmission time should be designed to be on the order of hundreds of microseconds while the current configuration in 4G is 1 ms. For this, enhancement in various areas of RAN such as packet/frame structure, modulation and coding schemes, new waveform designs, transmission techniques, and symbol detection need to be carried out. In order to reduce the delay in $T_{\text{Backhaul}}$, approaches such as advanced backhaul techniques, caching/fog enabled networks, and intelligent integration of AS and NAS can provide potential solutions. For $T_{\text{Core}}$, new core network consists of SDN, NFV, and various intelligent approaches can reduce the delay significantly. For $T_{\text{Transport}}$, MEC/fog enabled Internet/cloud/caching can provide reduced latency. 

In the following section, we discuss the constraints and approaches for achieving low latency.

\section{Constraints and Approaches for Achieving Low Latency}

There are major fundamental trade-offs between capacity, coverage, latency, reliability, and spectral efficiency in a wireless network. Due to these fundamental limits, if one metric is optimized for improvement, this may results in degradation of another metric. In the LTE system, the radio frame is 10 ms with the smallest TTI being 1~ms. This fixed frame structure depends on the modulation and coding schemes for adaptation of the transmission rate with constant control overhead. Since \emph{latency}  is associated with control overhead (cyclic prefix, transmission mode, and pilot symbols) which occupies a major portion of transmission time of a packet (approximately 0.3-0.4~ms per packet transmission), it is not wise to consider a packet with radio transmission time less than 1 ms. If we design a packet with time to transmit of 0.5 ms, more than $60 \%$ of the resources will  be used by control overhead \cite{EnablingLowLatency}. Moreover, retransmission per packet transmission takes around 8 ms, and removal of retransmission will affect packet error significantly. As a result, we need radical modifications and enhancements  in packet/frame structure and transmission strategy. In this regard: 

\begin{itemize}
\item First, a novel radio frame  reinforced by limited control overhead and smaller transmission time is necessary  to be designed. For reduction of control overhead, procedures for user scheduling, resource allocation, and channel training can be eliminated or merged.

\item Second, packet error probability for first transmission should be reduced with new waveforms and transmission techniques reducing the retransmission delay.

\item Third, since latency critical data needs to be dispatched immediately, techniques for priority of  data over normal data need to be identified. 

\item Fourth, synchronization and orthogonality are the indispensable aspects of OFDM that are major barriers for achieving low latency. Even though asynchronous mode of communication is  more favorable over synchronized operation in terms of latency, it requires additional spectrum and power resources \cite{nonrthogonalynchronous}. 

\item Fifth, since the latency for data transmission also depends on the delay between the core network and the BS, caching networks can be used to reduce latency by storing the popular data at the network edge.
\end{itemize}

 Researchers proposed various techniques/approaches for achieving low latency in 5G. As summarized in Fig. \ref{CatagorySolutions}, we divided the existing solutions into three major categories: (1) RAN solutions, (2) core network solutions, and (3) caching solutions. The RAN solutions include new/modified frame or packet structure, waveform designs, multiple access techniques, modulation and coding schemes, transmission schemes, control channels enhancements, low latency symbol detection, mmWave aggregation, cloud RAN, reinforcing QoS and QoE, energy-aware latency minimization, and location aware communication techniques. On the other hand, new entities such as SDN, NFV, MEC and fog network along with new backhaul based solutions have been proposed for the core network. The solutions of caching can be subdivided into caching placement, content delivery, centralized caching, and distributed caching, while backhaul solutions can be divided into general and mmWave backhaul. In the following sections, these solutions are described in further details.

\begin{table*}[!htbp]
 		\centering
 		\caption{\textsc{Overview of Techniques in RAN for Low Latency.}}
 		\label{Table_one}
 		\begin{tabular}{p{1.6cm}|p{1.8cm}|p{3.7 cm}|p{9.5cm}}
 			\hline \textbf{Case/Area}   &  \textbf{Reference} &  \textbf{Approach} & \textbf{Summary} \\ \hline \hline

 			 & \cite{7529226,7980747} & \textcolor{black}{Small packets/short TTI} & \textcolor{black}{Transmission of small scale data
 			 is investigated for packet loss rate of $10^{-9}$ and latency as low as 1 ms.} \\
 			
 			\cline{2-4} & \cite{GuanZRTBSK16} & Subcarrier spacing  & Subcarrier spacing is enlarged to shorten the OFDM symbol duration, and the number of OFDM symbols  is proposed to keep unchanged in each  subframe. \\ 
 			\cline{2-4} \textcolor {black}{Frame/Packet structure} & \cite{6881163,7504504,7432148,5Gradio} & Flexible OFDMA based TDD subframe   &\textcolor{black}{ TDD numerology is optimized for dense deployment with smaller cell sizes and larger bandwidth in the higher carrier frequencies.} \\
 			
 			\cline{2-4}  & \cite{TDDframe} &  Modification of physical subframe   & Different control and data part patterns for consecutive subframes, TX and RX control parts are proposed to be separated from each other, and from the data symbols with a GP, leading to total number of 3 GPs per subframe.\\
 			\cline{2-4} & \cite{7794610,pocovi2016impact,abreu2017a,soret2014fundamental} & Numerology, flexible sub frame and resource allocation& \textcolor{black}{ Numerology and subframe structure are defined considering  diverse carrier frequencies and bandwidths to envision 5G including \emph{low latency}. Cyclic prefix, FFT size, subcarrier spacing, and sampling frequency  were expressed as the function of carrier frequency.}\\
 			\hline
 			Advanced multiple access/Waveform    & \cite{waveform},\cite{5Gwave},\cite{5753092}	  & Filtered CP-OFDM, UFMC and FBMC & \textcolor{black}{ UFMC outperforms over OFDM by about $10\%$ in case of both large and small packets. FBMC demonstrates better performance in case of transmitting long sequences; however, it suffers during the transmission of short bursts/frames.}
 			\\ \hline
 			
 			  &  \cite{codingComp,Polar5G}  & Polar coding & Based on simulation and field test, polar coding has been proposed for 5G, outperforming over turbo coding in case of small packet transmission.\\
 			\cline{2-4}& \cite{Turbocoding} &	Turbo decoding with combined sliding window algorithm and cross parallel window (CPW) algorithm & A highly-parallel architecture for the latency sensitive turbo decoding is proposed combining two  parallel algorithms: the traditional sliding window algorithm and cross parallel window (CPW) algorithm.\\
 			\cline{2-4}& \cite{IFFTdesign} & New IFFT design with butterfly operation & Input signal of  IFFT processor corresponding to guard band are assigned as null revealing the existence of numerous zeros (i.e., $0$). If the sequence of OFDM symbol data which enter the IFFT is adjusted, the memory depth can be reduced from 1024 to 176.\\
 			\cline{2-4}Modulation and coding & \cite{DSSR} & Sparse code multiple access (SCMA) & \textcolor{black}{A dynamic shrunk square searching (DSSS) algorithm is proposed, which cuts off unnecessary communication control port (CCP) calculation along with utilization of both the noise characteristic and  state space structure.}\\
 			\cline{2-4} & \cite{7835946} & Priority to latency critical data & A latency reduction approach by introducing TDM of higher priority ultra-low latency data over other less time critical services is proposed which maps higher priority user data during the beginning of a subframe followed by the normal data. \\
 			\cline{2-4} & \cite{7403056} & Balanced truncation & \textcolor{black}{Balanced truncation is applied for the model reduction in the linear systems that are being coupled over arbitrary graphs under communication \emph{latency} constraints.}  \\
 			\cline{2-4} & \cite{ostman2016low} & Finite block length bounds and coding  & Recent advances in  finite-block length information theory are utilized in order to demonstrate  optimal design for wireless systems under strict constraints such as low latency and high reliability.\\
 			\hline
 & \cite{Asymmetricwindow} & Asymmetric window  & Asymmetric window is proposed instead of  well-known symmetric windows for reduction of cyclic prefix by $30\%$. This technique suppresses OOB  emission but makes the system more susceptible to channel induced ISI and ICI.  \\ 
		 			\cline{2-4}  & \cite{SheYQ16} & Transmission power optimization  & Transmission power is optimized by steepest descent algorithm considering transmission delay, error probability and queuing delay. \\ 
		 			  \cline{2-4} \textcolor {black}{Transmitter adaptation}& \cite{pathpacketredandency3} & Path-switching method and a packet-recovery method & Low latency packet transport system with a quick path-switching method and a packet-recovery method are introduced for a multi-radio-access technology (multi-RAT) environment.\\
		 			\cline{2-4} & \cite{RadioAccces} & Diversity & \textcolor{black}{Diversity could be employed through various approaches such as spatial diversity, time diversity, and frequency diversity.} \\
		 		\hline 		
			& \cite{sparsecontrol} & Control channel sparse encoding (CCSE) & CCSE is introduced in order to provide the control information using non-orthogonal spreading sequences.\\
		 		\cline{2-4} & \cite{Controlchannel} & Scaled control channel design & A scaled-LTE frame structure is proposed assuming the scaling factor to be 5 with a dedicated UL CCHs for all sporadic-traffic users in each transmission time interval  with possible smallest SR size. \\
		 		\cline{2-4} & \cite { UPLink } & Symbol-level frequency hopping and sequence-based sPUCCH & \textcolor{black}{A sequence-based sPUCCH (SS-PUCCH) incorporating two \textcolor{black}{SC-FDMA} symbols is introduced in order  to meet a  strict latency requirement. Symbol-level frequency hopping technique is employed to achieve frequency diversity gain and reliability enhancement.}\\ 
		 		\cline{2-4} Control signaling & \cite{7849058} & Radio bearer and S1 bearer management  & \textcolor{black}{Establishment of radio bearer and S1 bearer in parallel are proposed  where eNB and mobility management element (MME) manages and controls radio bearer and S1 bearer, respectively. The eNB sends only single control signal in order to configure radio bearers such as SRB1, SRB2 and DRBs, that decreases the signaling interaction rounds between the UE and the eNBs.}\\
		 		\cline{2-4} & \cite{ohseki2016fast} & Outer-loop link adaptation (OLLA) scheme& The proposed scheme controls the size of the compensation in the estimated SINR based on the time elapsed after a UE transits from an idle state to an active state, which helps to reduce latency for small packet applications.\\ \hline

 		\end{tabular}
 	\end{table*}
 	
 \begin{table*}[!htbp]
\ContinuedFloat
 	\centering
 	\caption{\textsc{Overview of Techniques in RAN for Low Latency (continued).}}
 	\label{Table_RAn_2}
 	\begin{tabular}{p{1.6cm}|p{1.8cm}|p{3.7 cm}|p{9.5cm}}
 		\hline \textbf{Case/Area}   &  \textbf{Reference} &  \textbf{Approach} & \textbf{Summary} \\ 
 		\hline \hline

 		  &  \cite{Improvinglatencyreliability}  & SM-MIMO detection scheme with ZF and MRC-ZF  &	A low-complexity and low latency massive SM-MIMO detection scheme  is introduced and validated using SDR platforms. The low complexity detection scheme is proposed with a combination of ZF and MRC-ZF.\\
 		\cline{2-4} & \cite{PacketStructure} & Linear MMSE & \textcolor{black}{A linear \textcolor{black}{MMSE} receiver is presented for low latency wireless communications using ultra-small packets.}\\ 
 		\cline{2-4} \textcolor {black}{Symbol detection} & \cite{GFDM3} & Space-time encoding and widely linear estimator  & \textcolor{black}{Space-time encoding is introduced within a GFDM block for maintaining overall low latency in the system. On the other hand, a widely linear estimator is used to decode the GFDM block at the receiver end,  which yields significant improvements in gain over earlier works.}\\ 
 		\cline{2-4} & \cite{compressedsensing,compressedsensing2, MASSIVEMTC}& Compressed sensing & Compressed sensing has been proposed to be effective in reducing latency of networked control systems if the state vector can be assumed to be sparse in some representation.\\
 		\cline{2-4} & \cite{7386169}& Low complexity receiver design & A low complexity receiver is designed and using this, the performance of an SCMA system is verified via simulations and real-time prototyping. This approach triples the whole system throughput while maintaining low latency similar to flexible orthogonal transmissions.\\
 		\hline
 		mmWave & \cite{DuttaMFZRZ15,FordZMDRZ16,Radio1,7794604} & mmWave based air interface  & Physical layer air
 		interface is proposed using mmWave aggregation. Large bandwidth along with various approaches such as small frame structure, mmWave backhaul and beamforming can help to achieve low latency.  \\ \hline
 	\textcolor {black}{Location aware communication} & \cite{5GETDLA,6891105,6924849} & Location information  & \textcolor{black}{Issues and
 		research challenges of 5G are discussed followed by the conclusion that  5G networks can exploit the location information and accomplish performance gains in terms of throughput and latency.}
 		\\ \hline
 QoS/QoE Differentiation & \cite{QoS1,wen20135g,QoSmmW2,QoS3, QS5, QoS6, QoS7,QoS8,QoS10,QoS11}	& Parameter manipulation & Differentiation of constraints on QoS and QoE can maintain
 low latency in 5G services including ultra high definition and
 3D video content, real time gaming, and neurosurgery.\\ \hline	
 Cloud RAN (CRAN)  & \cite{coudRAN1, CloudRan2, gao2015cloudlets, cloudRAN3, VPAAN} & Cloud architecture based RAN & CRANs combine baseband processing units of a group of base stations into a central server while retaining radio front end at the cell sides. Proper optimization of resources can ensure low latency along with capital expenditure reduction. \\ 
\hline
\end{tabular}
 \end{table*}

 \begin{figure*} [!htbp]
  	\centering
  	\subfigure[] {\label{fig:a}\includegraphics[width= 0.8\linewidth]{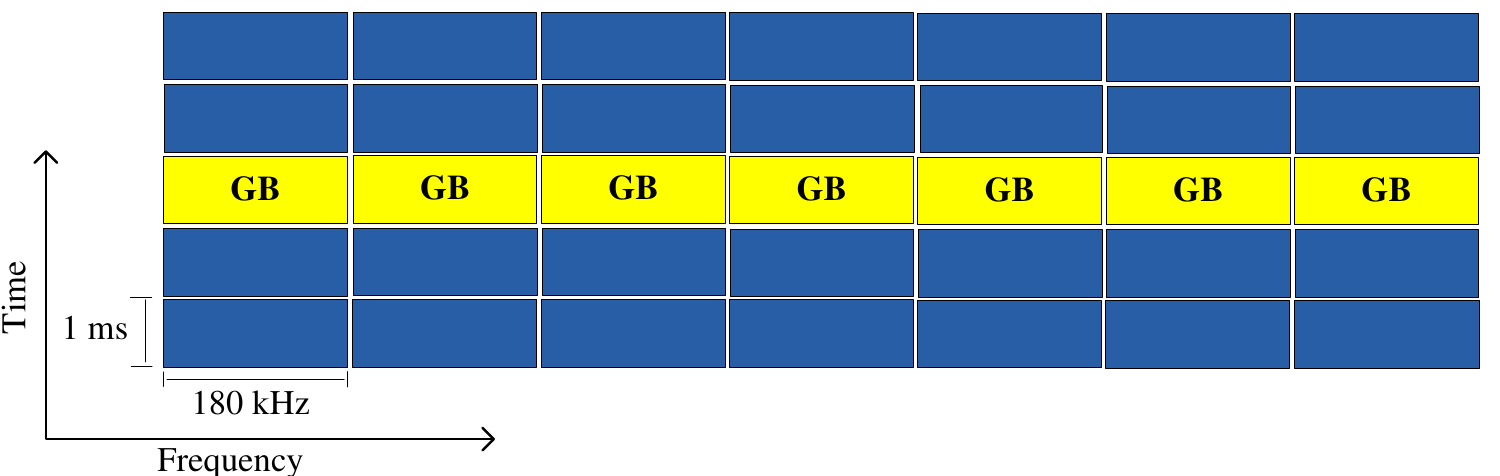}}
  	\subfigure[] {\label{fig:b}\includegraphics[width= 0.8\linewidth]{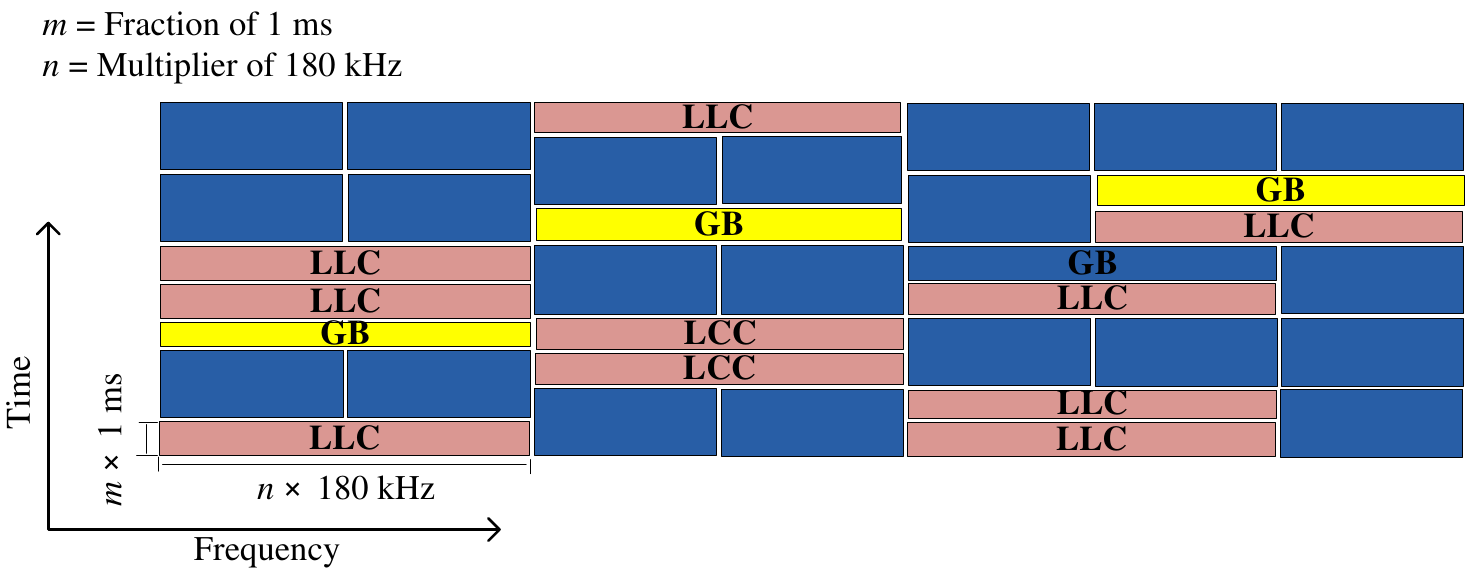}}	
  	\caption{Physical air interface (a) Conventional LTE radio frame, (b) Exemplary 5G radio frame with flexible time and frequency division for low latency \cite{5GETDLA} (GB: guard band; LLC: low latency communication).}
  	\label{fig:radioframe}
  \end{figure*}	
 	
\section{RAN Solutions for Low Latency}

To achieve \emph{low latency}, various enhancements in the RAN have been proposed. Referring to Table \ref{Table_one}, RAN solutions/enhancements include frame/packet structure, advanced multiple access techniques/waveform designs, modulation and coding scheme, diversity and antenna gain, control channel, symbol detection, energy-aware \emph{latency}  minimization, carrier aggregation in mmWave, reinforcing QoS and QoE, cloud RAN and location aware communication. In what follows, the detailed overview for each of these solutions is presented.

\subsection{Frame/packet structure}

In the RAN solutions, modification in the physical air interface has been considered as an attractive choice. In particular, most of the proposed solutions are on the physical (PHY) and medium access control (MAC) layers.

In LTE cellular network, the duration of a radio frame is 10 ms. Each frame is partitioned into 10 subframes of size 1 ms which is further divided into 0.5 ms units that are referred as a resource block (RB). Each RB spans 0.5 ms (6 or 7 OFDM symbols) in time domain and 180 KHz (12 consecutive subcarriers, each of which 15 KHz) in frequency domain. Based on this, the subcarrier spacing $\Delta f$
is 15 KHz, the OFDM symbol duration $T_{\mathrm{OFDM}}$ is $\frac{1}{\Delta f}=66.67\mu$s, the FFT size is 2048, the sampling rate $f_\mathrm{s}$ is $\Delta f \times N_{\mathrm{FFT}}=33.72$~MHz and the sampling interval $T_\mathrm{s}$ is $1/f_\mathrm{s}$.

To reduce TTI for achieving \emph{low latency}, the subcarrier spacing $\Delta f$ can be changed to 30 KHz \cite{GuanZRTBSK16}. \textcolor{black}{ This results the corresponding OFDM symbol duration $T_{\mathrm{OFDM}}$ to be 33.33~$\mu$s} and the FFT size $N_{\mathrm{FFT}}$ to become 1024 while sampling rate $f_\mathrm{s}$ is kept 30.72 MHz similar to LTE systems. The frame duration $T_s$=10 ms can be divided into 40 subframes in which each subframe duration $T_{\mathrm{sf}}$ is 0.25 ms and contains 6 or 7 symbols. Two types of cyclix prefixs (CPs) can be employed in this configuration with durations

\begin{equation}
T_{\mathrm{cp1}}=5/64\times N_{\mathrm{IFFT}}\times T_\mathrm{s} \approx 2.604~ \mu\mathrm{s},
\end{equation} 
\begin{equation}
T_{\mathrm{cp2}}=4/64\times N_{\mathrm{IFFT}}\times T_\mathrm{s} \approx 2.083 ~\mu\mathrm{s}.
\end{equation} 

\noindent A conventional LTE radio frame with equal sized RB and an exemplary 5G physical air frame are illustrated in Fig.~\ref{fig:radioframe}(a) and Fig.~\ref{fig:radioframe}(b), respectively.

 
In \cite{7529226}, an extensive analysis of the theoretical principles that regulates the transmission of small-scale packets with \emph{low latency}  and high reliability is presented with metrics to assess their performance. The authors emphasize  control overhead optimization for short packet transmission. In \cite{pocovi2016impact}, a flexible 5G radio frame structure is introduced in which the TTI size is configurable in accordance with the requirement of specific services. At low offered load, 0.25 ms TTI is an attractive choice for achieving \emph{low latency}  due to low control overhead. However, for more load, control overhead increases which affects reliability and packet recovery mechanism resulting in increased latency. This study argues to employ user scheduling with different TTI sizes in the future 5G networks. \textcolor{black}{In \cite{7980747}, the authors try to improve the outage capacity of URLLC and satisfy the low latency requirement of 5G using an efficient HARQ implementation with shortened transmission TTI and RTT. Moreover, some simulations are conducted in order to provide insights on the fundamental trade-off between  the outage capacity, system bandwidth, and the latency requirement for URLLC.}

In \cite{7794610}, the numerology and subframe structure are defined considering  diverse carrier frequencies and bandwidths for low latency 5G networks. Cyclic prefix, FFT size, subcarrier spacing, and sampling frequency  were expressed as a function of the carrier frequency. In \cite{5Gradio}, software defined radio (SDR) platform based 5G system implementation with strict latency requirement is presented. The scalability of the proposed radio frame structure is validated with \textcolor{black}{E2E} latency less than 1 ms. In \cite{GuanZRTBSK16}, the proposed subcarrier spacing is enlarged to shorten the OFDM symbol duration, and the number of OFDM symbols in each subframe is kept unchanged in the new frame structure for TDD downlink. The subcarrier spacing  is changed to 30 KHz resulting the corresponding OFDM symbol duration T $=33.33~\mu s$. The fast Fourier transform (FFT) size $N$ is  1024, while the sampling rate $f_{\mathrm {s}}$  is kept same as $30.72$~MHz. The frame duration $T_{\mathrm {s}}$ is still 10 ms with  40 subframes. 
 
  In \cite{TDDframe}, in order to have fully flexible allocations of different control and data RB in the consecutive subframes, TX and RX control RBs are proposed to be separated from each other and also from the data RB by  guard periods (GPs). This leads to total number of 3 GPs per subframe which separates them. Assuming symmetrical TX and RX control parts with $N_{\mathrm{ctrl\_s}}$ symbols in each and defining that same subcarrier spacing is used for control and data planes, with $N_{\mathrm{data\_s}}$ being the number of data symbols and $T_{\mathrm{symbol}}$ being the length of an OFDM symbol, the subframe length $T_{\mathrm{sf}}$ can be determined as
 \begin{eqnarray}
 T_{\mathrm{sf}}=(2N_{\mathrm{ctrl\_s}}+N_{\mathrm{data\_s}}(T_{\mathrm{symbol}}+T_{\mathrm{CP}}))+2T_{\mathrm{GP}}.
 \end{eqnarray}

In \cite{6881163}, the fundamental limits and enablers for low air interface latency are discussed with  a proposed flexible OFDM based TDD physical subframe structure optimized for 5G local area (LA) environment. Furthermore, dense deployment with smaller cell sizes and larger bandwidth in the higher carrier frequencies are argued as notable enablers for air interface latency reduction. In \cite{7504504}, a new configurable 5G TDD frame design is presented, which allows flexible scheduling (resource allocation) for wide area scenarios. The radical trade-offs between capacity, coverage, and latency are discussed further with the goal of deriving a 5G air interface solution capable of providing \emph{low latency}, high reliability, massive connectivity, and enhanced throughput. Since achieving low latency comes at cost of lower spectral efficiency, the proposed solution of the study includes control mechanisms for user requirement, i.e. whether the link should be optimized for low latency or high throughput. 
  
\begin{table}[!bp]
      	\centering
      	\caption{\textsc{PHY and MAC Layer Based Radio Interface Solutions for Low Latency.}}
      	\label{framePacket}	
      	\begin{tabular}{>{\centering\arraybackslash}m{1.2 cm}|p{2.5cm}|p{1.5cm}|p{1.5cm}}
      		\hline \textbf{References} & \textbf {Approach/Area} &\textbf {PHY layer}  & \textbf {MAC layer}\\ \hline
      		\hline \cite{pocovi2016impact} & Short TTI & \CheckmarkBold & \CheckmarkBold\\ 
      		\hline \cite{7794610,7529226} & Numerology and sub frame structure & \CheckmarkBold & \\ 
      		\hline \cite{GuanZRTBSK16} & Subcarrier & \CheckmarkBold & \\
      		\hline \cite{6881163} & Flexible subframe & & \CheckmarkBold\\ 
      		\hline \cite{5Gradio} & Flexible subframe implementation with SDR platform & \CheckmarkBold & \CheckmarkBold\\ 
      		\hline \cite{TDDframe} & Allocation of control and data RB & \CheckmarkBold & \CheckmarkBold \\ 
      		\hline \cite{7504504} & Radio frame and scheduling & \CheckmarkBold & \CheckmarkBold \\ 
      		\hline \cite{7432148}& Flexible TTI and  multiplexing & \CheckmarkBold & \CheckmarkBold \\ 
      		\hline \cite{7980747} & Efficient HARQ implementation with shortened transmission TTI and RTT &\CheckmarkBold &  \\ 
      		\hline \cite{abreu2017a} & Reservation of resources & & \CheckmarkBold \\ \hline
      		\cite{ soret2014fundamental} & Calculating the fundamental trade-offs among  three KPIs &\CheckmarkBold &\CheckmarkBold\\ \hline 
      		
      	\end{tabular} 
      \end{table} 
      
 \begin{table*}[!bp]
         	\centering
         	\caption{\textsc{Proposed Multiple Access Techniques for 5G.}}
         	\label{MAcomp}	
         	\begin{tabular}{m{3 cm} |p{4.5 cm}|p{4 cm}|p{4 cm}} 
         		\hline \textbf{Cases} & \textbf {IDMA  } \cite{niroopan2012user, fricke2005interleave}&\textbf {SCMA} \cite{SCMA1}   & \textbf {GFDM}  \cite{GFDM3,GFMC11}\\ \hline
         		\hline \vspace{0.5cm} Fundamental concept/features & \vspace{-0.5 cm} \begin{itemize} \item Specific interleaving \item User segregation \item Iterative multiuser identification \end{itemize} & \vspace{-0.5 cm}  \begin{itemize} \item Multiple dimensional code word \item QAM spreading combination \end{itemize} & \vspace{-0.5 cm} \begin{itemize} \item Block frame consists of time slots and subcarriers \item Non-orthogonal \item FFT/IFFT implementation \end{itemize}\\   
         		
         		\hline Low complexity & \CheckmarkBold & \CheckmarkBold & \\ 
         		
         		\hline Flexibility (in case of covering CP-OFDM and SC-FDE) &  & & \CheckmarkBold\\ 
         		
         		\hline Low latency &  & \CheckmarkBold & \CheckmarkBold \\ \hline
         		
         	\end{tabular} 
         \end{table*}      
      
A 5G flexible frame structure in order to facilitate users with highly diversified service requirements is proposed in \cite{7432148}. Although, in-resource physical layer control signaling is the basis of this proposed radio frame, it allows the corresponding data transmission based on individual user requirements. For this, it incorporates adaptable multiplexing of users on a shared channel with dynamic adjustment of the TTI in accordance with the service requirements per link. This facilitates optimization of the fundamental trade-offs between latency, spectral efficiency, and reliability for each link and service flow. In \cite{abreu2017a}, a scheme that reserves resources for re-transmission for a group of ultra reliable low latency communication UEs is presented. The optimum dimensioning of groups and block error rate (BLER) target can reduce the probability of contention for the shared retransmission resources. Moreover, the unused resources can be utilized for non-grouped UEs resulting in overall efficiency enhancement. 

In \cite{ soret2014fundamental}, fundamental trade-offs among three KPIs (reliability, latency, and throughput) in a 4G network is characterized, and  an analytical framework  is derived. In cases where the theory  can not be extended via mathematical formulations due to complexity of scenario in hand, some guidelines are provided to make the problem tractable. In order to  improve the aforementioned trade-offs between these  KPIs in future 5G systems different candidate techniques are proposed.  
 
 The above approaches of frame/packet structure to achieve \emph{low latency}  at the RAN level are tabulated in Table \ref{framePacket}.

\subsection{Advanced Multiple Access Techniques/Waveform}

 \begin{figure}[t!]
  \centering 
  \includegraphics[width=1\linewidth]{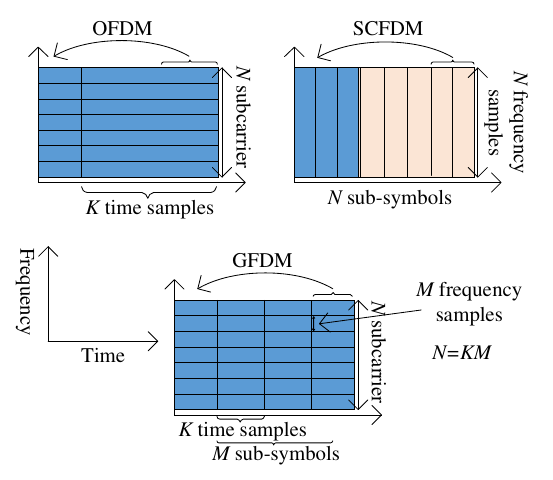}
  \caption{Slot placment in GFDM, OFDM and SCFDM.}
  \label{GFDM}
  \end{figure}
 
 Different kinds of candidate multiple access (MA) techniques and waveforms including orthogonal, non orthogonal and asynchronous  access have been proposed for \emph{low latency}  communication \cite{nonrthogonalynchronous, waveform, 5Gwave, 5753092}. Since synchronization and orthogonality (integral to OFDM) is a hindrance for achieving \emph{low latency}, asynchronized non orthogonal multiple access techniques have been discussed in \cite{nonrthogonalynchronous}. Reduction of symbol duration to $67~\mu$s is not a promising solution in critical time budgeting. In this regard, interleave division multiple access (IDMA) has been introduced in \cite{niroopan2012user, fricke2005interleave} for generating signal layers. The IDMA is a variant of the CDMA technique which uses specific interleaving for user segregation in lieu of using a spread sequence to the individual user. Here, channel coding, forward error correction coding and spreading are combined into a single block by a low rate encoder. The spreading can not be considered as a distinct and special task. Interleaving usually utilizes a simpler iterative multiuser identification approach. However, this approach needs further rigorous investigation.
 
 \begin{table*}[tp!]
 	\centering
 	\caption{\textsc{Waveform Contender for  5G.}}
 	\label{UMFC}	
 	\begin{tabular}{p{3.5cm}|p{4 cm}|p{4cm}|p{4cm}}
 		\hline \textbf{\hskip 30pt Cases} & \textbf {OFDM} &\textbf {FBMC} \cite{SCMA2,multcarrier} & \textbf {UFMC} \cite{UMFCcomp,GFMC11}\\  \hline
 		\hline \vskip 1pt Filtering & Generalized filtering to all subcarriers of entire band & Filtering to each subcarriers & Generalized filtering to a group of consecutive subcarriers\\ \hline  Requirement of coordination  &  Higher & Lower & Lower \\ \hline
 		\textcolor{black}{Time-frequency efficiency (due to CP and guard band)}& \vskip 1pt 0.84 & \vskip 1pt 1 &\vskip 1pt 1 \\
 		\hline
 		ICI (in case of lower degree of synchronization with UEs and eNBs)  & \vskip4pt  Higher   &\vskip4pt Lower  &\vskip4pt  Lower \\ \hline
 		 Performance & Performs well for large packets with well coordination & Performs well for large packets with less coordination & Performs well for short packets with less coordination\\ \hline
 		
 	\end{tabular} 
 \end{table*}
    
 In order to supply synchronization and orthogonality, sparse code multiple access (SCMA) and non orthogonal multiple access (NOMA) have been presented in \cite{SCMA1} for 5G scenarios. In SCMA, symbol mapping and spreading are combined together, and the mapping of multi dimensional codeword over incoming bits is performed directly from SCMA codebook. SCMA is comparatively simpler and has superior performance over low density version of CDMA. Another modulation technique that aims to reduce latency is referred as the  generalized frequency division multiplexing (GFDM) is introduced in \cite{GFDM3,GFMC11}. The flexibility of covering both the cyclic prefix OFDM (CP-OFDM) and single carrier frequency domain equalization (SC-FDE), and block structure of GFDM help to achieve \emph{low latency}. A typical mapping structure of GFDM, OFDM and SC-FDM is illustrated in Fig.~\ref{GFDM}. The overall comparison among IDMA, SCMA and GFDM is presented in Table \ref{MAcomp}.

Filter bank multi carrier (FBMC) has been a strong candidate waveform for 5G \cite{SCMA2,multcarrier}. FBMC demonstrates better performance in case of transmitting long sequences; however, it suffers during the transmission of short bursts/frames. For usage of cyclic prefix, wide frequency guards and more required coordination, OFDM may be inefficient in case of low latency communication \cite{GFMC11}. Universal filtered multi-carrier (UFMC) \cite{UMFCcomp,GFMC11} is upgraded version of FBMC which offsets the disadvantage of FBMC. It outperforms OFDM by about $10\%$ in cases of time frequency efficiency, inter carrier interference (ICI) and transmissions of long or short packets. Additionally, UFMC preforms better than FBMC  in the case of very short packets while demonstrating similar performance for long sequences. These make UFMC as the one of the best choices for next generation low latency communication.
 
In case of UFMC, the time domain transmit vector \cite{UMFCcomp} for a user is superposition of sub-band wise filtered components. The time domain transmit vector for a particular multi-carrier symbol of user  $k$ with filter length $L$ and FFT length $N$ is

\begin{equation}
\underset{(N+L-1)\times 1}{\mathbf{X_{\textit{k}}}}  =\sum_{i=1}^{B} \underset{(N+L-1)\times N}{\mathbf{F_{\textit{ik}}}} \vspace{10mm} \underset{ N \times n_i}{\mathbf {V_{\textit{ik}}}} \vspace{10mm} \underset{ n_i \times 1}{\mathbf{S_{\textit{ik}}}},
\vspace{-0.65 in}
\end{equation} 

\noindent where

\begin {itemize}

\item  $\mathbf{S}$ is the complex QAM symbol vector;

\item  $\mathbf{V}$ is the transformed time domain vector by IDFT matrix; In this case, the relevant columns of the inverse Fourier matrix are incorporated in accordance with the respective subband position within entire available band;

\item $i$ is the index of each subband of $B$;

\item $\mathbf{F}$ is a Toeplitz matrix. It is comprised of filter impulse response, and performs the linear convolution.

\end {itemize}

\textcolor{black}{The symbol duration of $(N+L-1)$ samples is determined by the filter length and  FFT size. Filtering per block per subcarrier allows spectrally broad filters in pass band and shorter in time domain compared to FBMC. The reduced time yields \textcolor{black}{shortened} OFDM CP. The filter ramp up and ramp down in \textcolor{black}{shorten} time domain ensures symbol shaping in a way that allows protection against ISI and robustness for multiple access users. Furthermore, being orthogonal with respect to complex plain, complex modulation symbol can be transmitted without further complication.}

Another advantage of UMFC is the ability of using different subcarrier spacings or filter times for users in different subbands. If a user  uses  FFT size $N_1$ and filter length $L_1$, and another user  uses filter length and FFT size of $N_2$ and $L_2$ respectively, then UFMC symbol duration can be designed such that $N_1+L_1-1= N_2+L_2-1$. This makes UFMC a remarkable adaptive  modulation scheme with capability to be tailored easily under various characteristics of communications, including delay/Doppler spread variations in the radio channel and user QoS needs. The comparative discussion among OFDM, FBMC and UMFC is presented in Table \ref{UMFC}. 

\subsection{Modulation and Channel Coding} 

\textcolor{black}{Although use of small packets is a potential approach for achieving \emph{low latency}, appropriate modulation and coding is required for small packet transmission for acceptable reliability.} In the literature, mainly three types of coding schemes are proposed for 5G. As presented in \cite{codingComp}, low-density parity-check (LDPC) and polar codes outperform turbo codes in terms of small packets while for medium and large packets, the opposite is true. While small packet is a requirement for \emph{low latency}, other aspects such as implementation complexity, performance in practical test, and flexibility need to be investigated. In \cite{Polar5G}, polar code has been tested in field for 5G considering various scenarios: air interface, frame structure, settings for large and small packets, OFDM, and filtered OFDM (f-OFDM) waveforms. In all cases, polar code performed better than turbo codes which makes it  a candidate channel coding scheme for 5G. The comparison among the schemes are illustrated in Table \ref{5Gchanenlcoding}.

\begin{table}[!b]
 	\centering
 	\caption{\textsc{Comparison among Channel Coding Schemes for Low Latency \cite{codingComp}.}}
 	\label{5Gchanenlcoding}	
 	\begin{tabular}{p{2cm}|p{1.2cm}|p{1cm}|p{1.6cm}|p{0.9cm}}
 		\hline \textbf{Cases} & \textbf{Turbo coding} \cite{Turbocoding} & \textbf {LDPC-PEG}\cite{codingComp} & \textbf {Convolutional coding}\cite{codingComp} & \textbf {Polar codes} \cite{Polar5G} \\ \hline
 		\hline  Algorithm complexity for coding 1/3 of 40 bits with respect to turbo codes& 100$\%$ & 98$\%$ & 66.7$\%$ & 1.5$\%$ \\
 		\hline Algorithm complexity for coding 1/3 of 200 bits with respect to turbo codes& 100$\%$& 98$\%$ & 66.7$\%$ & 110.7 $\%$\\
 		\hline Performance in short packets &  & \CheckmarkBold && \CheckmarkBold\\ 
 		\hline Performance in medium packets &  \CheckmarkBold & &\CheckmarkBold & \\ \hline
 	
 	\end{tabular} 
 \end{table}
  
  \textcolor{red}{
  \begin{table*}[!bp]
  \centering
  	\caption{\textsc{Overview of Solutions in Transmitter Adaptation for Low Latency.}}
  	\label{Transmission}	
  	\begin{tabular}{p{1.8 cm}|p{3.2 cm}|p{5.2 cm}|p{5.2 cm}}
  	\hline  \textbf{Reference} & \textbf{Techniques}&\textbf{\textcolor{black}{Merits}} &\textbf{\textcolor{black}{Demerits}}\\ \hline \hline
  	   \cite{Asymmetricwindow}  & Asymmetric window & \textcolor{black}{Reduces cyclic prefix by 30\% and  maintains good OOB suppression along with latency \textcolor{black}{reduction}.} & \textcolor{black}{It assumes that spectral mask will be \textcolor{black}{stricter} in 5G networks.}\\  \hline
  	   \cite{SheYQ16} & Transmission power optimization & \textcolor{black}{Queuing delay is \textcolor{black}{considered} in optimization along with transmission delay and packet error.} & \textcolor{black}{No uniform cross layer information exchange format is provided. Besides that, cross layer signaling may result extra overhead in the nodes.} \\ \hline
  		\cite{pathpacketredandency3} & Path-switching and packet-recovery method& \textcolor{black}{Provides fast switching and recovery method in multi-RAT environment.} & \textcolor{black}{It depends on availability of good channel for path switch, and packet recovery may affect the resiliency.}\\ \hline
  		\cite{RadioAccces} & Diversity gain & \textcolor{black}{Option of various diversity gains such as space, time and spatial gain for low latency transmission.}& \textcolor{black}{Gain depends on various aspects such as beam forming, beam training and antenna array.} \\  \hline
  		 \cite{mmwavephy}& Beam forming using mmWave& \textcolor{black}{Design and analysis of MAC layer under realistic conditions.} & \textcolor{black}{Proper channel model in mmWave is under development.}\\  \hline
  		 \cite{FD1,FD2,FD3,FD4,rahmati2015price, rahmati2017price} & Full duplex communication in same channel & \textcolor{black}{Improves throughput, \textcolor{black}{reduces} latency and upholds PHY layer security.} & \textcolor{black}{Crosstalk between the transmitter (Tx) and the receiver (Rx), internal interference, fading, and path loss.}\\ \hline
  	\end{tabular} 
  \end{table*}}

 In \cite{Turbocoding}, a highly-parallel architecture for the \emph{latency}  sensitive turbo decoding is proposed by combining two  parallel algorithms: the traditional sliding window algorithm and cross parallel window (CPW) algorithm. New IFFT design  with butterfly operation is  proposed in \cite{IFFTdesign}, which reduces IFFT output data delay through the reduction of IFFT memory size and butterfly operation (e.g. addition/subtraction). \textcolor{black}{ Input signal of the IFFT processor corresponding to guard band is assigned as zero (i.e. `0') revealing the existence of numerous zeros. } If the sequence of OFDM symbol data which enter the IFFT is adjusted, the memory depth can be reduced from 1024 to 176.
 
A dynamic shrunk square searching (DSSS) algorithm is proposed in \cite{DSSR}, which cuts off unnecessary communication control port (CCP) calculation by utilizing  both the noise characteristic and  state space structure. In this way, it can maintain close to optimal decoding performance in terms of the block error rate (BLER). This results in reduction of delay in communication. In \cite{7835946}, a \emph{latency}  reduction approach by introducing time division multiplexing (TDM) of higher priority ultra-low \emph{latency}  data over other less time critical services is proposed, which maps higher priority user data during the beginning of a subframe followed by the normal data. In \cite{7403056}, balanced truncation is applied for the model reduction in the linear systems that are being coupled over arbitrary graphs under communication \emph{latency} constraints. In \cite{ostman2016low}, recent advances in  finite-block length information theory are utilized in order to demonstrate  optimal design for wireless systems under strict constraints such as low latency and high reliability. For a given set of constraints such as bandwidth, latency, and reliability the bounds for the number of the bits that can be transmitted for an OFDM system is derived. 
 
 \subsection{\textcolor{black}{Transmitter Adaptation}}

A representative set of approaches for reducing \emph{latency}  using  transmission side processing  are tabulated in Table \ref{Transmission}, which will be overviewed in the rest of this subsection.

In  \cite{Asymmetricwindow}, an asymmetric window is proposed instead of  well-known symmetric windows for reduction of cyclic prefix by $30\%$, and hence reducing \emph{latency}  due to reduced overhead. This technique suppresses out of bound (OOB)  emission but makes the system more susceptible to channel induced inter symbol interference (ISI) and inter carrier interference (ICI). Transmission power optimization by the steepest descent algorithm considering transmission delay, error probability and queuing delay is proposed in \cite{SheYQ16}. In \cite{pathpacketredandency3}, low \emph{latency}  packet transport system with a quick path-switching and a packet-recovery method is introduced for a multi-radio-access technology (multi-RAT) environment. In \cite{RadioAccces}, use of  diversity gain is proposed as a solution for capacity enhancement and latency reduction. Diversity could be achieved through various approaches such as spatial diversity, time diversity, and frequency diversity. 
%
\begin{table*}[!tp]
		\centering
	\caption{\textsc{Overview of Solutions in Control Signaling for Low Latency.}}
	\label{Controlchannel}	
	\begin{tabular}{p{1.5 cm}|p{3.5 cm}|p{5.2 cm}|p{5.2 cm}}
	\hline  \textbf{References} &\textbf{Techniques}&\textbf{\textcolor{black}{Merits}} &\textbf{\textcolor{black}{Demerits}}\\ \hline  \hline
	      \cite{sparsecontrol} & Control channel sparse encoding (CCSE)& \textcolor{black}{Uses non-orthogonal spreading sequences.} & \textcolor{black}{Needs field test for further validation.}\\ \hline
		\cite{Controlchannel} & Dedicated UL CCHs & \textcolor{black}{Provides CCH for sporadic packets with small size scheduling request (SR).} & \textcolor{black}{Requires dedicated CCH in each TTI and well designed scheduling request (SR) detector at BS. Also \textcolor{black}{considered} scenario with UL and DL signal space of 10 and 40 bits, \textcolor{black}{spatial diversity of 16}, and bandwidth 10~MHz may not be always feasible.} \\ \hline
	  \cite{Uplinkframe}& Sub slotted data and control channel & \textcolor{black}{Two symbols are used in each subslot which is compatible with current LTE.} & \textcolor{black}{Reliability issue is not addressed.} \\ \hline
	  \cite{UPLink }& SS-PUCCH consists of SC-FDMA symbol & \textcolor{black}{More robust to channel fading \textcolor{black}{compared} to reference signal based PUCCH. Symbol level frequency hoping harnesses frequency diversity gain with enhanced reliability.}& \textcolor{black}{Need to be validated by field test.}\\ \hline
		\cite{7849058} & Radio bearer and S1 bearer management & \textcolor{black}{Control overhead and latency are decreased for both light and heavy traffic. It ensures 100\% accessibility of all UEs.} & \textcolor{black}{The technique is more suitable for very large traffic networks such as vehicle networks.}\\ \hline
		\cite{ohseki2016fast}& Outer-loop link adaptation & \textcolor{black}{Besides \textcolor{black}{reduction} of latency for small packets, it can boost throughput just after changing  from idle state to connected state.} & \textcolor{black}{Needs field test for further validation.} \\ \hline
		\cite{SlotRadio}& Slotted TTI based radio resource management & \textcolor{black}{It can be implemented as an extension of LTE-A.} & \textcolor{black}{Validation through simulation and field test is not presented.}\\ \hline
		\cite{RLC001}& Adaptive radio link control (RLC) & \textcolor{black}{Besides latency \textcolor{black}{reduction}, it improves throughput and \textcolor{black}{reduces} processing power.} & \textcolor{black}{Control and data plane need to be separated.} \\ \hline
		\cite{7464864} & SDN based control plane optimization & \textcolor{black}{Using bandwidth rebating strategy, balance between cost and performance is maintained.} & \textcolor{black}{For large number of players (vehicles), the game can be complicated. Also real world field test is required for performance evaluation.}\\ \hline
		\cite{8031628} & SDN based X2 signaling management & \textcolor{black}{It reduces signaling overhead and handover latency.} & \textcolor{black}{The approach has been investigated for only femtocells.}\\ 
		\hline
		\cite{8108504} & Inter BS data forwarding and make-over-handover&\textcolor{black}{This technique reduces X2 communication, processing and reconfiguration delays.} & \textcolor{black}{Cases such as packet loss, handover failure, scenario with poor communication link are not considered for performance evaluation.} \\  \hline 
		\cite{8064563} & Optical connected splitters with dynamic bandwidth allocation allocation and tailored MAC protocol & \textcolor{black}{It ensures X2 latency less than 1~ms.} & \textcolor{black}{Needs field test for further validation.}\\ \hline
	\end{tabular} 
\end{table*}

In \cite{mmwavephy}, a mmWave based switched architecture system is proposed  where control signals use low-resolution digital beamforming (to enable multiplexing of small control packets) with analog beamforming in the data plane (to enable higher order modulation). This reduces the overhead  significantly due to the control signaling which results in more resources for data transmission. This technique leads to reduction of round trip  \emph{latency}  in the physical layer.

Recent advancements in full duplex (FD) communication comes forward with feature of doubling the capacity, improving the feedback, and latency mechanism meanwhile upholding steady physical layer security \cite{FD1,FD2,FD3,FD4,rahmati2015price, rahmati2017price}. Various proposed techniques of 5G networks such as massive MIMO and  beamforming technology providing  reduced spatial domain interference can be contributive for FD realization \cite{FD2}. Besides that intelligent scheduling of throughput/delay critical packets along with proper rate adaption and power assignment can results in capacity gain and reduction of \emph{latency}. However, this field needs to be extensively investigated  for studying capacity and latency trade offs.

\subsection{\textcolor{black}{Control Signaling}} 

When the packet size is reduced  as envisioned in 5G systems, control overhead takes the major portion of the packet. Addressing this, various approaches are proposed in order to reduce the control channel  overhead. The potential solutions targeting the control channel enhancements to achieve low \emph{latency}  are illustrated in Table~\ref{Controlchannel}.    


In \cite{sparsecontrol}, control channel sparse encoding (CCSE) is introduced with vision to \textcolor{black}{transmit the control information by means of non-orthogonal spreading sequences.} A scaled-LTE frame structure is proposed in \cite{Controlchannel} assuming the scaling factor to be 5 with  dedicated UL~ control channels (CCHs) for all sporadic-traffic users in each TTI  with possible smallest scheduling request (SR) size. In \cite{Uplinkframe}, short TTI based uplink frame has been proposed for achieving E2E \emph{latency}  no longer than 1~ms. In the proposed scheme, subslot consisting of 2 symbols has been proposed for uplink data and control channel. A sequence-based sPUCCH (SS-PUCCH) incorporating two single carrier-frequency division multiple access (SC-FDMA) symbols is introduced in \cite{UPLink } in order  to meet a  strict latency requirement. Symbol-level frequency hopping technique is employed to achieve frequency diversity gain and reliability enhancement.

In the proposed procedure of \cite{7849058}, establishment of radio bearer and S1 bearer in parallel are proposed  where eNB and mobility management element (MME) manages and controls radio bearer and S1 bearer, respectively. The eNB sends only single control signal in order to configure radio bearers such as SRB1, SRB2 and DRBs, that decreases the signaling interaction rounds between the UE and the eNBs. In \cite{ohseki2016fast}, a new  outer-loop link adaptation (OLLA) scheme is proposed. The scheme controls the size of the compensation in the estimated SINR based on the time elapsed after a UE transits from an idle state to an active state, which helps to reduce \emph{latency}  for small packet applications. The study \cite{SlotRadio} proposed a slotted TTI based radio resource management for LTE-A and 5G in order to achieve low latency. The approach can serve low latency services utilizing short TTI and enhance download control channel (ePDCCH). 

\begin{table*}[!tp]
\centering
	\caption{\textsc{Overview of Solutions in Symbol Detection for Low Latency.}}
	\label{Detection}	
	\begin{tabular}{p{1.5 cm}|p{3.5 cm}|p{5.2 cm}|p{5.2 cm}}
		\hline \textbf{Reference} &\textbf{Technique/Approach}&\textbf{\textcolor{black}{Merits}} &\textbf{\textcolor{black}{Demerits}}\\ \hline \hline
		\cite{Improvinglatencyreliability} &SM-MIMO detection with ZF and MRC-ZF& \textcolor{black}{Significant improvement of SINR and latency is achieved compared to other schemes. Also the method is validated in live environment designed by SDR platform.}& \textcolor{black}{Availability of large number of antennas is assumed.}\\ \hline
		\cite{PacketStructure}&Linear MMSE receiver& \textcolor{black}{Reduces latency along with throughput gain improvement.} & \textcolor{black}{\textcolor{black}{ It is not clear  how much latency can be reduced in this scheme.}}\\ \hline
		\cite{GFDM3} &Space-time encoding and widely linear estimator& \textcolor{black}{Significant improvements in terms of symbol error rate and latency over earlier works.}& \textcolor{black}{Complexity at receiver side is increased.}\\ \hline
		\cite{compressedsensing,compressedsensing2, MASSIVEMTC}& Compressed sensing & \textcolor{black}{CS algorithm exhibits reduced complexity and increases
			reliability. \textcolor{black}{It} is compatible with the current LTE systems \textcolor{black}{as well, and} requires less measurement (resource) to decode control information. It provides sub-Nyquist sampling method for reconstruction of sparse signal efficiently in a linear system.}& \textcolor{black}{It is challenging to design CS and sparse recovery system considering diverse wireless conditions and input conditions.}\\ \hline
		\cite{7386169} &Low complexity receiver in SCMA system.& \textcolor{black}{The prototype triples capacity while maintaining low \textcolor{black}{latency.}} & \textcolor{black}{More suitable for MTC. Needs field test for further validation.}\\
		\hline
	\end{tabular} 
\end{table*}
The study \cite{RLC001} proposed a novel mechanism  that introduces an adaptive radio link control (RLC) mode which dynamically alternates between unacknowledgment mode (UM) and acknowledgment mode (AM) according to the real-time analysis of radio conditions. This technique reduces  system \emph{latency}  and processing power, and improves throughput using UM. On the other hand, it improves data reliability  by activating AM during the degraded radio conditions. In \cite{7464864}, SDN based control plane optimizing strategy is presented to balance the \emph{latency}  requirement of vehicular ad hoc network (VANET), and the cost on radio networks. The interaction between vehicles and controller is formulated and analyzed as a two-stage Stackelberg game followed by optimal rebating strategy, which provides reduced \emph{latency} compared to other control plane structures.

\textcolor{black}{In \cite{8031628}, \textcolor{black}{SDN-based} local mobility management with \textcolor{black}{X2} forwarding is proposed where total handover  signaling is minimized by reduction of inter node signaling exchanges and \textcolor{black}{X2} signaling forwarded to centralized SDN system. This approach can reduce handover latency while reducing of signal overhead. In \cite{8108504}, QoS, CQI and other parameter based data utilization is proposed among eNBs to reduce X2 latency, processing and reconfiguration delays. Additionally, make-before-handover is proposed for low latency 5G services for no data interruption. In order to meet stringent latency in X2 interface, enhanced passive optical network (PON) based radio network is proposed in \cite{8064563}, where the splitters are connected thorough optical connections. Following this, dynamic bandwidth allocation algorithm and tailored MAC protocol are introduced for achieving less than 1~ms latency over X2 interface.}


\subsection{Symbol Detection} 
\begin{figure}[t!]
	\centering 
	\includegraphics[width=1\linewidth]{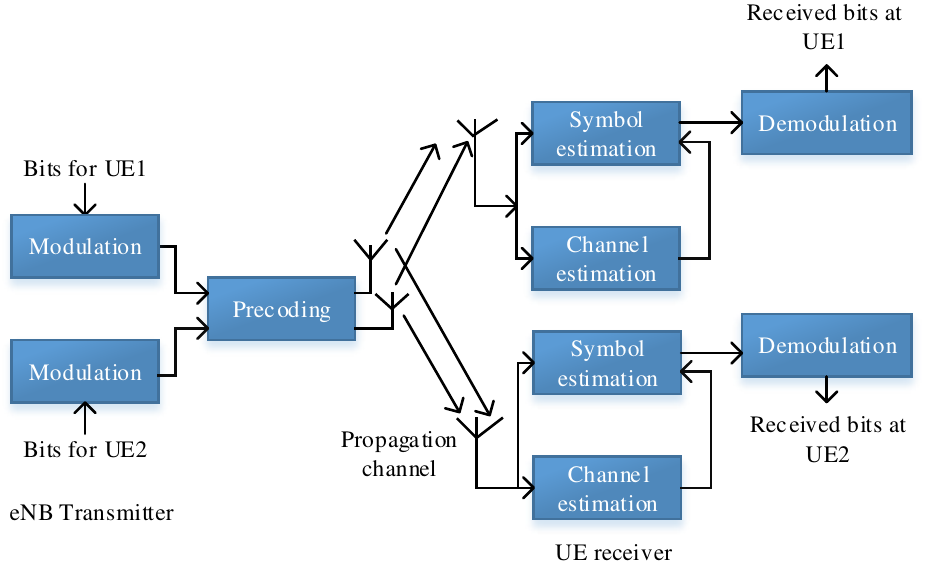}
	\caption{Transmission and symbol detection in cellular network.}
	\label{MIMO}
\end{figure}

\textcolor{black}{As illustrated in  Fig. \ref{MIMO}, symbol detection encompasses various processes such as channel estimation and decoding, which can all contribute into the overall \emph{latency}. The related literature in the symbol detection side for  latency reduction are tabulated in Table \ref{Detection}.}

\begin{table*}[!hbp]
	\centering
	\caption{\textsc{Overview of Solutions in mmWave Communications for Low Latency.}}
	\label{Table_mmWave}	
	\begin{tabular}{p{1.2 cm}|p{9 cm}|p{1.5 cm}|p{1.5 cm}|p{1.5 cm}}
		\hline  \textbf{References} &\textbf{Techniques}&\textbf{\textcolor{black}{PHY layer}} &\textbf{\textcolor{black}{MAC layer}} &\textbf{\textcolor{black}{NET layer}}\\ \hline \hline
		\cite{DuttaMFZRZ15} & mmWave based MAC layer frame structure& &\CheckmarkBold &\\ \hline
		\cite{FordZMDRZ16} & Low latency core network architecture, flexible MAC layer, and congestion control & &\CheckmarkBold &\CheckmarkBold\\ \hline
		\cite{Radio1} & mmWave based physical layer air interface with basic numerology and logical channel arrangement &\CheckmarkBold  & &\\ \hline
		\cite{7794604} & Low latency frame structure with beam tracking &\CheckmarkBold  & &\\ 
		\hline
	\end{tabular} 
\end{table*}

In \cite{Improvinglatencyreliability}, a low-complexity and low-latency massive SM-MIMO detection scheme is introduced and validated using SDR platforms. The low complexity detection scheme is proposed with a combination of zero forcing (ZF) and maximum-ratio-combining-zero-forcing (MRC-ZF). In \cite{PacketStructure}, a linear minimum mean square error (MMSE) receiver is presented for low latency wireless communications using ultra-small packets. The estimation of receiver filter using the received samples is proposed during the data transmission period in lieu of  interference training period. Additionally, soft decision-directed channel estimation is argued using the data symbols for re-estimation of the channels. In \cite{GFDM3}, space-time encoding is introduced within a GFDM block in order to achieve transmit diversity for overall low latency in the system. On the other hand, a widely linear estimator is used to decode the GFDM block at the receiver end,  which yields significant improvements in terms of symbol error rate and \emph{latency} over earlier works.

In \cite{compressedsensing,compressedsensing2, MASSIVEMTC}, compressed sensing is proposed for latency reduction in networked control systems if the state vector can be modeled as  sparse in some representation domain. In \cite{7386169}, a low complexity receiver design is proposed and the superiority of an SCMA system  is verified via simulations. In addition, it is demonstrated with a  real-time prototype that the whole system throughput triples while maintaining low latency similar to flexible orthogonal transmissions.

\subsection{mmWave Communications} 


Carrier aggregation using the mmWave spectrum is widely considered to be a promising candidate technology for 5G, capable of providing massive bandwidth and ultra low \emph{latency}. The mmWave technology is especially critical for VR/AR type of applications which require high throughput and low \emph{latency}. The works in mmWave spectrum for achieving low latency is summarized in Table~\ref{Table_mmWave}. 

\begin{table*}[!tp]
	\centering
	\caption{\textsc{Location-Aware Communications for Low Latency.}}
	\label{Tablelocation2}	
	\begin{tabular}{p{1.8 cm}|p{3 cm}|p{5 cm}|p{5 cm}}
		\hline  \textbf{References} &\textbf{Techniques}&\textbf{\textcolor{black}{Merits}} &\textbf{\textcolor{black}{Demerits}}\\ \hline\hline
		\cite{6924849} & Location information utilization in protocol stack & \textcolor{black}{Latency, scalability and robustness can be improved.}& \textcolor{black}{Location accuracy, spatial channel modeling, balancing trade-off between location information and channel quality metric are challenging.}  \\ \hline
		\cite{5GETDLA} & Physical layer parameters design using FFT, frame duration and local area (LA) physical channel & \textcolor{black}{Improves spectral and energy efficiency along latency reduction.} & \textcolor{black}{Needs field test for further validation.}\\ \hline
		\cite{6891105} & Utilization of channel quality and traffic statistics from small cell & \textcolor{black}{Coexistence capability  with overlay LTE-A network, sleeping modes, contention based data channel, channel quality indicator and interference statistics.} & \textcolor{black}{The technique is more feasible for small cells.}\\ 
		\hline
	\end{tabular} 
\end{table*}
In \cite{DuttaMFZRZ15}, a new frame design for mmWave MAC layer  is introduced which provides several improvements including adaptable and smaller transmission intervals, dynamic locations for control signals, and the capability of  directional multiplexing for control signals (dynamic HARQ placement). It addresses ultra low latency along with the multiple users, short bursty traffic and beam forming architecture constraints. The study \cite{FordZMDRZ16} focuses on three critical higher-layer design areas: low latency core network architecture, flexible MAC layer, and congestion control. Possible solutions to achieve improvements in these critical design areas are short symbol periods, flexible TTI, low-power digital beam forming for control, and low latency mmWave MAC, which can all be considered for  data channel, downlink control channel, and uplink control channel. 

In \cite{Radio1}, in order to decrease the latency of the system, two different physical layer numerologies are proposed. The first approach  is  applicable for indoor or line of sight (LOS) communications, and the second one is suitable for non line of sight (NLOS) communications. This is justified by some  channel measurements experiments in $28-73$~GHz range.  In \cite{7794604}, a 5G mmWPoC system is employed to evaluate the throughput functionality in field tests at up to 20 km/h mobile speed in an outdoor LOS environment. Additionally, some improvements for a frame design is  obtained which decrease the \emph{latency}  in the field tests. In the experiments, it is observed that the new slotted frame design can decrease the RTT to 3~ms for $70\%-80\%$ of the cases in experiments, alongside the  observed throughput up to 1~Gbps.

\subsection{Location-Aware Communications for 5G Networks}

Location knowledge (in particular, the communication link distance) can be considered as a criterion of received power, interference level, and link quality in a wireless network. Therefore,  overhead and delays can be reduced with location-aware resource allocation techniques because of the possibility  of channel quality prediction
	beyond traditional time scales. The literature on location-aware communications regarding low latency are tabulated  in Table \ref{Tablelocation2}.

In \cite{6924849}, several approaches are presented for monolithic location aware 5G devices  in order to identify  corresponding signal processing challenges, and describe  how location data should be employed across the protocol stack from a big picture perspective. Moreover, this work also
presents several open challenges and research directions that
should be solved before 5G technologies employ mmWave to
achieve the performance gains in terms of latency, connectivity
and throughput. In \cite{5GETDLA}, 5G flexible TDD is proposed  for  local area (5GETLA) radio interface with FFT size of 256 and 512, and short frame structure to achieve \emph{latency}  lower than 1~ms. The packets of size of less than 50 kbits can be transmitted with E2E \emph{latency}  of 0.25~ms. The main focus in designing physical layer parameters is on FFT, frame duration and physical channel (LA). 

In \cite{6891105}, a novel numerology and radio interface architecture is presented for  local area system by flexible TDD, and  frame design. \textcolor{black}{The proposed framework ensures coexistence  with overlay LTE-A network, sleeping modes, contention based data channel, and channel quality indicator and interference statistics. }Here, the channel quality and traffic statistics are accumulated from the small cells which can help to gain high throughput and low \emph{latency}. Especially, in order to reduce the \emph{latency}, the delay due to  packets containing critical data for  the higher layer protocols, for instance transmission control protocol (TCP) acknowledgment (ACK)
	packets, must  be optimized. To do so, one possible approach is to carry out the retransmissions as quick as possible compared  to the
	higher layer timers. Moreover, capability of  data transmission to a contention
	based data channel (CBDCH) can play a key role here. As a result,
	by introducing CBDCH in small cells that are not highly loaded, the average latency of  small packets transmission can
	be decreased considerably.

\subsection{\textcolor{black}{QoS/QoE Differentiation}}

\begin{table}[!bp]
	\centering
	\caption{\textsc{Literature Overview Related to QoS/QoE Differentiation.}}
	\label{Table_QoSQoE}
	\begin{tabular}{p{1 cm}|p{4.5cm}|p{0.5cm}|p{0.5cm}}
		\hline \textbf{Reference} & \textbf{Techniques/Approaches} & \textbf{\textcolor{black}{QoS}} & \textbf{\textcolor{black}{QoE}}\\ \hline\hline
		\cite{QoS1} & mmWave utilization with beam tracking & \CheckmarkBold & \CheckmarkBold\\  \hline
		\cite{wen20135g}& SDN and cloud technology & \CheckmarkBold & \CheckmarkBold \\  \hline
	    \cite{QoSmmW2}&  QoS-aware multimedia scheduling & \CheckmarkBold &\\  \hline
		\cite{QoS3}& Client based QoS monitoring architecture & \CheckmarkBold&\\  \hline
		\cite{QS5}& Colored conflict graph & \CheckmarkBold &\\  \hline
		\cite{QoS6}& QoS architecture with heterogeneous statistical delay bound & \CheckmarkBold &\\  \hline
		\cite{QoS7}& Dynamic energy efficient bandwidth allocation scheme & \CheckmarkBold &\\ \hline
		\cite{QoS8} & Predictive model based on Internet video download &  &\CheckmarkBold \\ \hline
	    \cite{QoS10} & Routing using proximity information & &\CheckmarkBold \\ \hline
	   \cite{QoS11}& Predictive model based on empirical observations & &\CheckmarkBold \\ \hline
	\end{tabular}
\end{table}

\textcolor{black}{Differentiation of constraints on QoS and QoE can maintain low \emph{latency}  in 5G services including ultra high definition and 3D video content, real time gaming, and  neurosurgery.} The related literature  on QoS and QoE control for low latency services  are tabulated in Table \ref{Table_QoSQoE}. 

Abundance of mmWave bandwidth and extensive use of beamforming techniques in 5G will allow high QoS and QoE overcoming the  resource and sharing constraints \cite{QoS1}. However, current  transmission protocols and technologies cannot be employed simply  for addressing technical issues in 5G. The mapping of diverse services including latency critical service to the optimal frequency, SDN and cloud technologies can ensure to achieve  the best QoS and QoE, as discussed in \cite{wen20135g}. In \cite{QoSmmW2}, a QoS-aware multimedia scheduling approach is proposed using  propagation analysis and proper countermeasure methods to meet the QoS requirements in the mmWave communications. Mean opinion score (MOS) which is a criterion for user satisfaction can be employed for  functionality evaluation of the newly  presented QoS approach and well-known distortion driven scheduling in different frequency ranges. 

Client based QoS monitoring architecture is proposed in \cite{QoS3} to address the issue of QoS monitoring from server point of view. Different criteria such as  bandwidth, error rate and signal strength  are proposed  with the well-known RTT delay for maintaining desirable QoS.  A colored conflict graph is introduced in \cite{QS5} to capture multiple interference and QoS aware approaches in order to take the advantage of beamforming antennas. In this case, reduction in call blocking and handoff failure helps to have a better QoS for multi class traffic. Each device can be sensitive to time based on its application. This can be considered as an issue for  QoS provisioning. To address this, a  novel QoS architecture is presented in \cite{QoS6} with heterogeneous statistical delay bound over a wireless coupling channel. The authors presented the dynamic energy efficient bandwidth allocation schemes in \cite{QoS7}, which improve system quality significantly and maintain QoS.

Previous QoS criteria which consist of  packet loss rate, network latency, peak signal-to-noise ratio (SNR)  and RTT are not sufficient for streaming media on Internet, and therefore,  users' perceived satisfaction (i.e. QoE) needs to be addressed \cite{QoS8,QoS9Survey}. Higher QoS may not ensure the satisfactory QoE. Different routing approaches of video streams in the mobile network operators’ scenario is discussed in \cite{QoS10}  for substantial refinement in QoE considering bit rate streams, low jitter, reduced startup delay and smoother playback.   A predictive model from empirical observations is presented in \cite{QoS11} to  address interdependency formulated as a machine learning problem. Apart from that, a predictive model of user QoE for Internet video is proposed in \cite{QoS8}.

\subsection{\textcolor{black}{CRAN and Other Aspects}}

\begin{figure}[!tp]
	\centering 
	\includegraphics[width=1\linewidth]{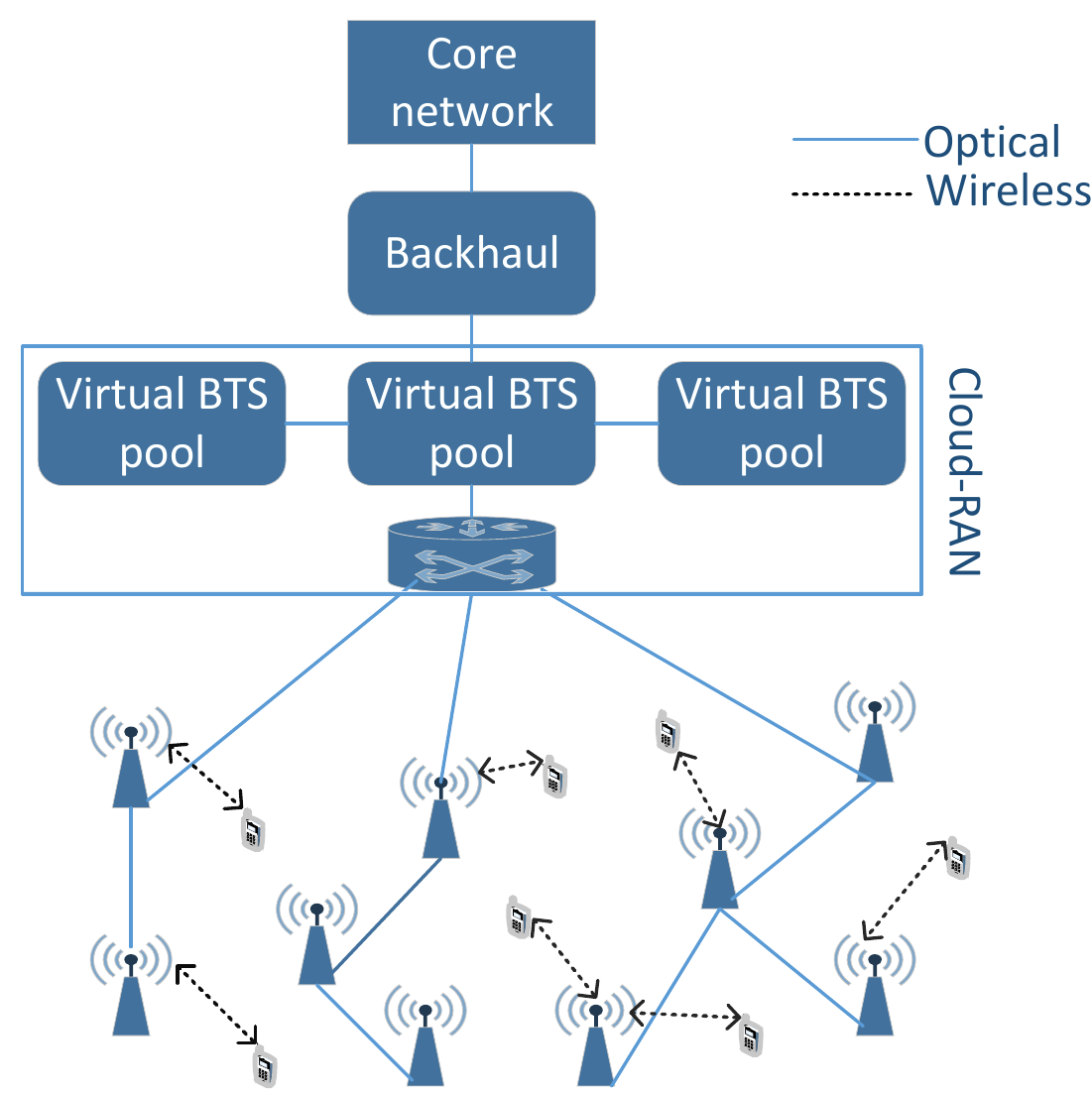}
	\caption{Cloud-RAN architecture  in 5G networks.}
	\label{cloudRAN1}
\end{figure}

\textcolor{black}{Cloud radio access network (CRAN) (as illustrated in Fig.~\ref{cloudRAN1}) is introduced for 5G in order to reduce the capital expenditure (CAPEX)  and simplify the network management~\cite{NEMNEM1973}.} CRANs combine baseband processing units of a group of base stations into a central server retaining radio front end at the cell sides. However, this requires connection links with delay of $250~\mu$s to support 5G low latency services. In order to meet strict latency requirements in CRAN,  two optimization techniques including (i) fine-tuned real-time kernel for processing latency and (ii) docker with data plane development kit (DPDK) for networking latency have been proposed in \cite{coudRAN1}. The experimental results clearly demonstrate the effectiveness of the approaches for latency optimization. In \cite{CloudRan2}, split of PHY and MAC layer in a CRAN with Ethernet fronthaul is proposed, and verified  through experimental test followed by latency interpretation. It is found that \emph{latency}  for packets of size 70 and 982 bytes is $107.32~\mu$s and $128.18~\mu$s confirming $20\%$ latency increase from small to large packet. The promising results affirm that  latency critical services in 5G can be supported by CRAN. {\color{black}In \cite{gao2015cloudlets}, it is demonstrated \textcolor{black}{ that based on experimental results} from Wi-Fi
	and 4G LTE networks, offloading the traffic to cloudlets
	outperforms the  response times by 51\% in comparison to cloud
	offloading. }

The CRAN can utilize backhaul information to redistribute users for QoE maximization and adaptation of temporal backhaul constraints. Corresponding to this, authors in \cite{cloudRAN3} proposed a centralized optimization scheme to control the cell range extension offset so as to minimize the average network packet delay. In \cite{VPAAN}, a CRAN on the basis of the optical network (PON) architecture is presented which is called virtualized-CRAN (V-CRAN). The proposed scheme can dynamically affiliate each radio unit (RU) to a digital unit (DU) which results in coordination of multiple RUs with their corresponing DU. Moreover, definition of virtualized BS (V-BS) is brought up which is able to  mutually send shared signals from several RUs to a user. V-CRAN can reduce \emph{latency}  for joint transmission due to the following reasons. First, it can provide  more than enough bandwidth for data transmission
	between RUs and DU. Furthermore,  for each DU, a
	dedicated hardware/software is assigned that can be utilized by joint transmission controller in order to provide data and signaling for RUs. The last but not least, in order to handle the load distribution between DUs, the virtualized PON can connect a
	DU directly via a linecard (LC).

Though mmWave will be \textcolor{black}{the  major} contributor in attaining 5G goals, spectrum below 6~GHz is always the primary choice due to less attenuation, supporting long distance, and antenna compatibility. \textcolor{black}{Moreover, conventional cellular networks are usually deployed within the  expensive licensed bands
		and they use reliable core networks that are optimized to provide low-volume delay-sensitive services  such  as  voice. However, with appearance of  high-volume
		delay-insensitive resource-hungry applications   including multimedia downloads, such conventional networks may not be cost-effective anymore  \cite{hong2014cognitive,ding2015limits}. Such concerns can be tackled or at least partially addressed by spectrum sharing in 5G network improving spectrum and energy efficiency along with QoS/QoE control \cite{8088528,7564899, spectrum1, spectrum2, spectrum3, spectrum4}. In \cite{7147778}, in order to exploit the TV white space for  D2D communications underlying existing cellular infrastructure, a framework is proposed. A  location-specific TV white space database is proposed in which D2D service can be provided using a look-up table for the D2D link so that it can    determine its maximum permitted emission power  in the unlicensed digital TV band to avoid interference.} In \cite{7368824}, \textcolor{black}{ a} QoE driven dynamic and intelligent spectrum assignment scheme is proposed which can support both cell and device level spectrum allocation. This technique enhances not only the spectrum utilization, but also can maintain \textcolor{black}{desired} QoE including \emph{latency} aspect. The optimization problem of network sum rate and access rate with  resource allocation and QoS constraints in D2D communication is presented in \cite{7581088}. To solve this, a fast heuristic  algorithm is proposed to reduce computational complexity resulting desired QoS such as latency. In \cite{7878689}, \textcolor{black}{a} game theory and interference graph based optimization problem considering user scheduling, power allocation and spectrum access is presented with an aim to maximize user satisfaction across the network. Two algorithms including spatial adaptive play iterative (SAPI) learning are proposed to achieve Nash equilibrium. In \cite{7368822}, \textcolor{black}{a} multi agents  based cognitive radio framework is proposed for effective utilization of spectrum and meet the goals of 5G including latency. Here, sensing capability of secondary users (SU) is \textcolor{black}{replaced} by \textcolor{black}{spectrum agents (SA)} where \textcolor{black}{the} user can switch between SU mode and SA mode based on available spectrum information.
\begin{figure*} [!t]
	\centering
	\subfigure[] { \label{fig:a}\includegraphics[width= 0.47\linewidth,height=5.7 cm]{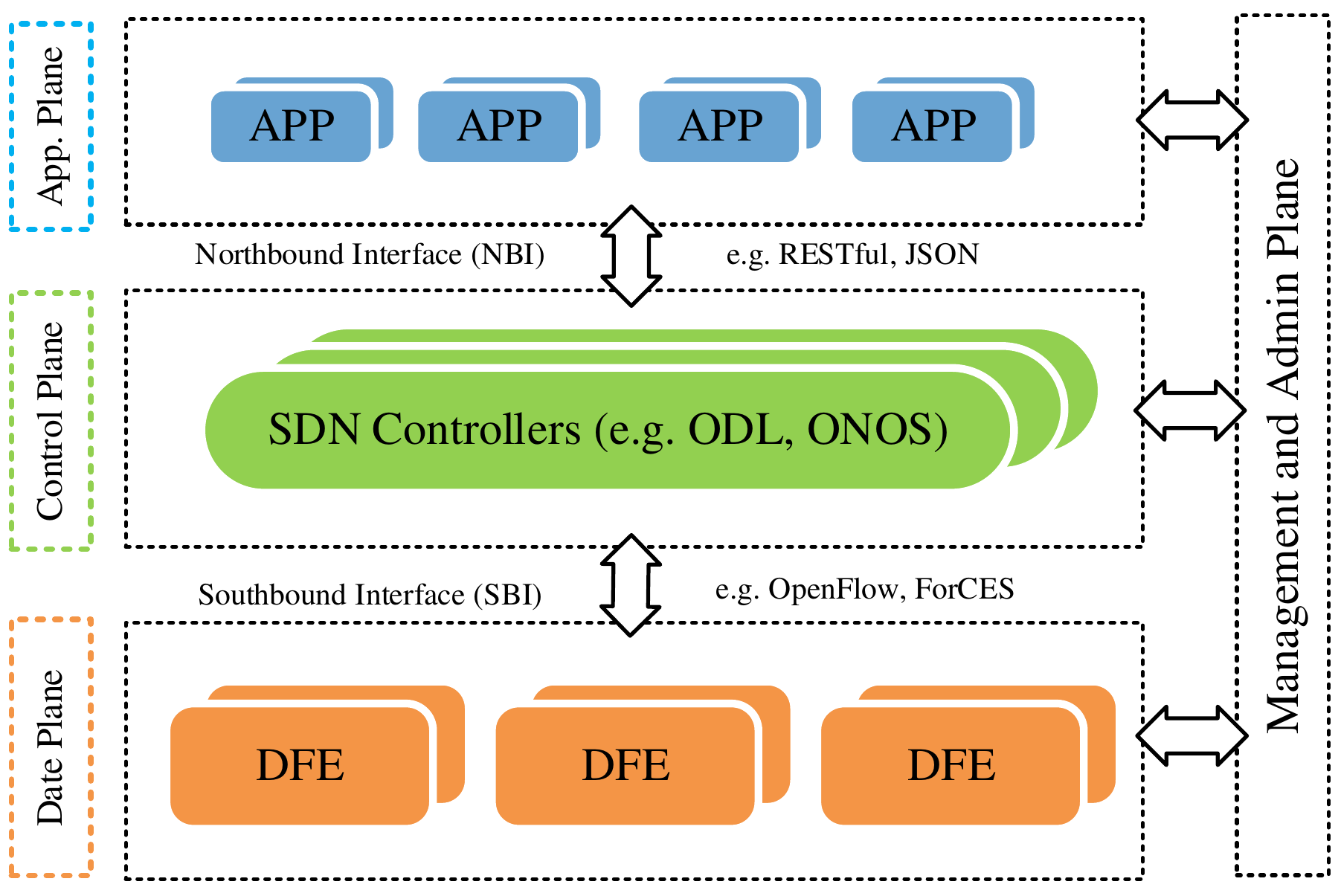}}
	\hspace*{2mm}
	\subfigure[] {\label{fig:b}\includegraphics[width= 0.47\linewidth,height=6 cm]{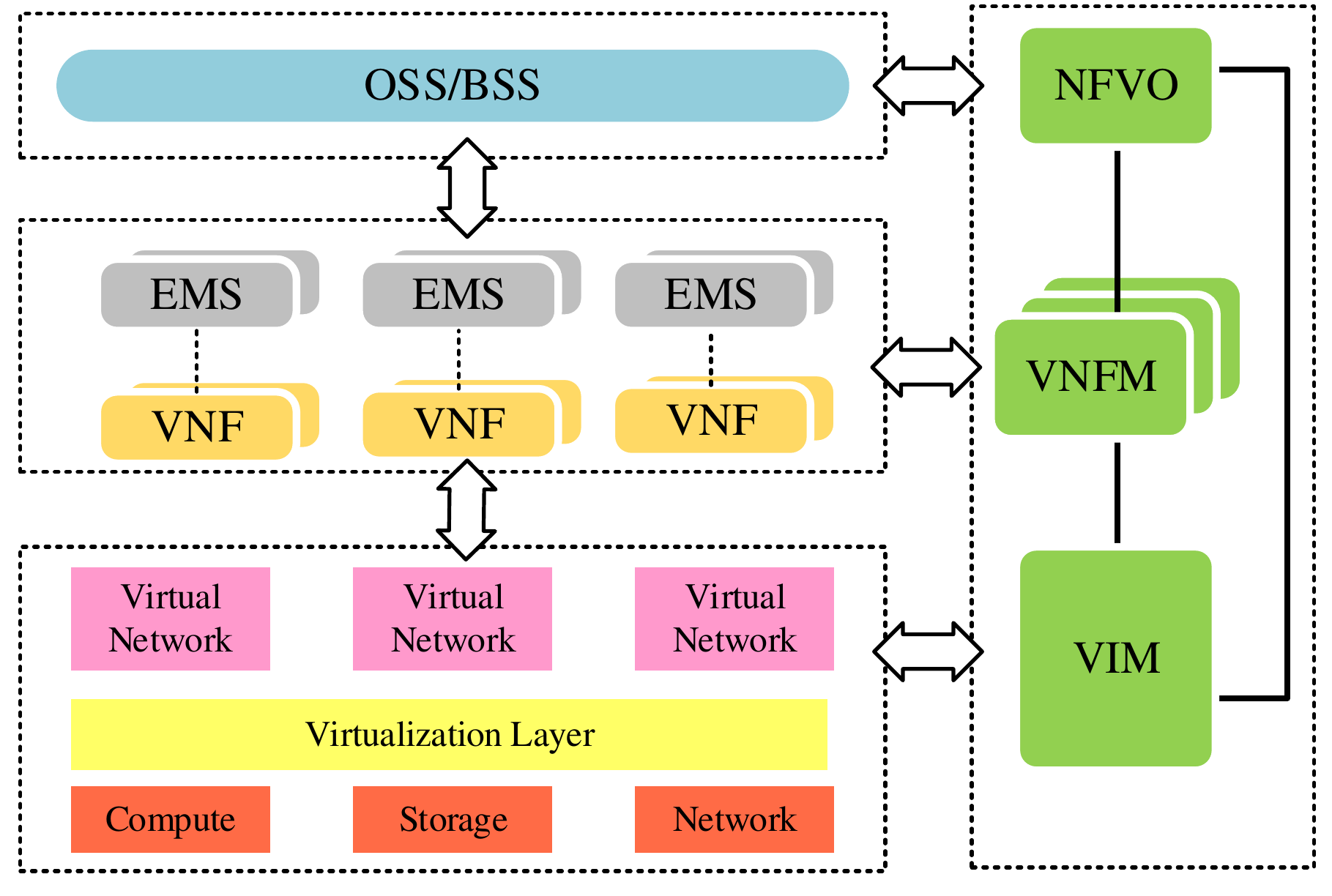}}	
	\caption{Simplified example architectures for core network (a) Architecture of SDN \cite{SDNref}, and (b) Architecture of NFV \cite{NFVAR} (APP: application; ODL: opendaylight platform; ONOS: open network operating system; DFE: dyna forming engineering; OSS: operations support systems; BSS: base station subsystem; EMS: element management system;  NFVO: NFV orchestrator; VNFM: virtual network function manager; VIM: virtualized infrastructure manager).}
	\label{SDNFV}
\end{figure*}

\begin{table*}[!bp]
	\centering
	\caption{\textsc{Overview of Techniques in Core Network for Low Latency.}}
	\label{Tablecore}
	\begin{tabular}{c|p{3cm}|p{3cm}|p{7cm}}
		\hline \textbf{Case/Area}   &  \textbf{Reference} &  \textbf{Approach} & \textbf{Summary} \\ 
		\hline  \hline
		
		& \cite{EnablingLowLatency}, \cite{SDNAr,SDN5G,SDNtactile,heinonen2014dynamic}, \cite{SDNFV,7010528,costa20175g, wang2017efficient, ford2017provisioning,marquezan2016understanding,moradi2014softmow,moradi2014softmow2,thyagaturu2016sdn,jain2016comparison } &  SDN-based  architecture  &	The architecture of 5G is proposed based on SDN with 5G vision to meet large throughput, massive connectivity and low latency.\\ 
		\cline{2-4}Core Network Architecture &   \cite{SDNFV,NFV,cau2016efficient,LatencyawareNFV,mobiltymnage,costa20175g, wang2017efficient,ford2017provisioning, taleb2014toward,jain2016comparison,6963800} & NFV-based architecture & NFV invalidates the dependency on  hardware platform and makes  easy deployment of  EPC functions as well as the sharing of resources in the RAN. This can reduce the E2E \emph{latency}  with improved throughput performance.\\  
		\cline{2-4} & \cite{{EnablingLowLatency}, mobiltymnage,NFV,LatencyawareNFV, wang2017survey, ford2017provisioning} & MEC/fog-based network & MEC/fog provides computation and storage near user end and also separates the data plan from control plan. These reduces the \emph{latency} .\\
		\hline
	\end{tabular}
\end{table*}
Before starting the packet transmission in data applications,  a tolerable initial delay can be considered. Thus, this short delay can be employed to  decrease the energy required to operate the small cell base stations (SBSs). With the goal of   average power consumption reduction of SBSs, when the load is low, under-utilized SBSs can be  are switched off. As a result, the users and the network can be able to save energy by postponing  transmissions. By doing so, the users have time to wait for an SBS with better link quality. It can be observed that sleeping mode operation for energy efficiency improvements in SBSs will also introduce a source of \emph{latency}. In \cite{celebi2013energy}, the energy-efficiency versus delay trade-off is investigated and optimality conditions for UE’s transmit power is derived.  By postponing the access of the users, energy efficiency of the system can be improved. The optimal threshold distance  is derived in order to minimize the  average distance between a user and an SBS. Simulation results demonstrate that the energy consumption of SBS can be reduced by about 35\% if some of the SBSs are switched off.

In \cite{hhhg}, authors  introduced an  energy efficient, low-complexity technique for load-based sleep mode optimization in densely deployed 5G small cell networks.  By defining a  new analytic model, the distribution of the traffic load of a small cell  is  characterized using Gamma distribution. It is shown that  the network throughput can be improved significantly while some amount of energy is saved by taking the benefit of the initial delay. In \cite{7752490}, impact of average sleeping time of BS and association
radius on the mean delay in an UDN is investigated using a M/G/1/N queuing model. An explicit equation of delay is derived and the effect of average sleeping time and association radius on the mean delay is analyzed.

\textcolor{black}{In \cite{7959596}, the remote PHY (R-PHY) and the remote PHYMAC (R-PHYMAC) based modular broadcast cable network is proposed for the access network. In this architecture,  R-PHYMAC can achieve lower mean upstream packet delay compared to R-PHY for bursty traffic and long distance over 100 miles. In \cite{rakib2015virtual}, a virtual converged cable access platform (CCAP) system and procedure is proposed for hybrid fiber coaxial (HFC) cable network. In this method, {\color{black}a} new digital optical configuration is introduced to receive data packets with capability to convert {\color{black}them} into RF waveforms. This method improves space and power {\color{black}requirements} while enhancing operational flexibility. In \cite{DBLP:journals/corr/abs-1708-08902}, a novel remote FFT (R-FFT) module is proposed which can perform physical layer processing of FFT module towards RF transmission. This module reduces fronthaul bit-rate requirement for CRAN while providing solution for unified data over cable service interface specification (DOCSIS) and LTE service over HFC cable network.}

\section{Core Network Solutions for low latency}

\textcolor{black}{To meet the vision of 5G encompassing ultra low \emph{latency}   in addition to enhancements in the RAN, drastic changes  are also proposed in the core network. The new core network includes some new entities such as SDN, MEC, and NFV as well as new backhaul techniques \cite{SDNFVsurvey,MECSurvey,7503119}.} These enhancements aim to  reduce the processing time, bypass several protocol layers, and ensure seamless operation. The core network solutions for low \emph{latency} are reviewed in further detail in the rest of this section.

\subsection{Core Network Entities}

\textcolor{black}{The SDN and NFV are assumed to be the main candidates for the design of 5G core network \cite{agyapong2014design}. Based on this, in this section, we mainly focus on the role of SDN and NFV technologies in latency reduction in 5G core network.} Exemplary architectures of SDN and NFV of the 5G core network are illustrated in Fig. \ref{SDNFV}.  The existing literature  on the core network entities that can facilitate to  achieve low \emph{latency}  are summarized in Table \ref{Tablecore}.

\textcolor{black}{The EPC which is developed by 3GPP for the LTE cellular network has some limitations which affect the latency of the overall system. One concern is that the control plane and the data plane in the EPC are not fully separated. There is a level of coupling  between Serving Gateway (SGW) and Packet Data Network Gateway (PGW). Decoupling of control plane and data plane seems necessary because they have different network QoS criteria to be met. In particular, the control plane needs low latency to process signaling messages, while the data plane requires high throughput to process the data. Thus, in order to design such planes  efficiently, it is preferable to  decouple them completely. Based on the literature, SDN and NFV can be employed in EPC architecture in order to decouple data plane and control plane and have a seamless operation of core network functions \cite{brunstrom2017sdn,kasgari2018stochastic}.}\\

\textcolor{black}{After modification of NFV  based EPC in which the whole network elements are implemented using softwares running on Virtual Machines (VM), control plane and user plane can be separated by employing SDN in EPC. An SDN controller can act as \textcolor{black}{an interface} between the decoupled planes. In addition to several advantages of SDN/NFV-based user plane and control plane separation, including independent scalability, flexibility of flow distribution, and better user mobility management, such a decoupling can have considerable effect on reducing the {\em{latency}} as well. This plane decoupling can facilitate the mobile edge computing technology which decreases the latency. However, adding an SDN controller to the network can be another source of the latency for the system. On the other hand, the scalability of SDN controller can be addressed by deploying several controllers. Thus, there is a trade off here between the scalability of controllers and latency increase which should be considered in the design process for specific applications \cite{7010528}.}

\textcolor{black}{
	Another limitation is that the data plane of the LTE EPC is implemented in a centralized manner. Even the users that need to communicate locally have to transmit their traffic in a hierarchal system ending with few number of centralized PGWs which increases the E2E latency. Although centralized implementation of network can facilitate the management and monitoring the network by operators, it increases the E2E latency which can not meet the applications requiring low latency including autonomous driving, smart-grid or automated factory. Thus, this kind of implementation leads to inefficient system performance and high latency which can not meet the 5G vision \cite{wang2017survey}.}

\textcolor{black}{Recently, by emergence of the new technologies such as  cloud computing, fog networks, mobile edge computing, NFV and SDN, the implementation of the network can be more distributed \cite{costa20175g, wang2017efficient }. By employing such technologies, the CAPEX and OPEX of the network can be reduced considerably. Moreover, by bringing the elements of core network closer to the users, the \textcolor{black}{E2E} delay can be decreased significantly.} \textcolor{black}{ In \cite{ford2017provisioning}, authors proposed  SDN/NFV-based
	MEC networks algorithms that can enable
	the data plane to create a distributed MEC by placement of network functions at a  distributed
	manner. They demonstrated that the proposed scheme can reduce the redundant \textcolor{black}{{data center}} capacity around 75\%, and meet the 5G latency requirement along with considerable backhaul link bandwidth reduction.  }

\textcolor{black}{Mobility  management
	in core network based on SDN can potentially introduce some delays.
	In  \cite{marquezan2016understanding}, the main contributers for processing delays in an SDN-based mobility management system is discussed. By implementing two proactive and reactive solutions for mobility management using Mininet and OpenFlow, it is observed that with high probability (almost 95\%) in the proactive mobility management 
	system, the overall processing latency is around  the median value.  By visualizing all of EPC entities as decentralized VMs in different locations, in 
	\cite{taleb2014toward}, a carrier cloud architecture is introduced. To improve the \textcolor{black}{E2E} latency of the users, the concept of Follow-Me-Cloud is presented. The main point of this concept is that all parts of the network can keep track of the movement of the user which results in seamless connectivity and lower \textcolor{black}{E2E} latency. In \cite{moradi2014softmow}, \cite{moradi2014softmow2}, the authors proposed a decentralized scheme for control plane called SoftMow  which is  a hierarchical 
	reconfigurable network-wide control plane.  The proposed control plane includes  geographically  distributed  controllers  where each controller is responsible to serve the network in its particular location. The number of the
	levels in this hierarchical scheme can be designed based on the  available latency budgets. } \\


In \cite{EnablingLowLatency}, authors presented  an LTE compliant architecture to decrease \emph{delay} for combination of fog networks, MEC and SDN in which the architecture is supposed to take the advantage of NFV in the evolved packet core (EPC) functions. Following this, optimization of general packet radio service (GPRS) tunneling protocol (GTP) is introduced for supporting low latency services.  GTP tunnels management is accomplished by a novel element between the eNB and the mobile network interface with the Internet. In \cite{SDNAr,SDN5G}, SDN is proposed along with some changes in the existing 4G architecture for moving forward towards 5G. The changes include  reduction of number of serving gateways (S-GW) and elimination of  some protocol layers. In SDN based system, virtualization is possible, and routes can be optimized. This will allow handling of QoS by setting specific rules in the switches along the data path. Network coding integrated with SDN is proposed in \cite{SDNtactile} for low latency and reduced packet-retransmission. Network coding can work as network router and can be integrated with SDN, which provides seamless network operation and reduction in \emph{latency}.

The NFV is proposed as a major entity of 5G core network in \cite{SDNFV, NFV,cau2016efficient,mobiltymnage,LatencyawareNFV}. NFV removes the dependency on the hardware platform and makes flexible deployment of  EPC functions as well as sharing of resources in RAN. This can reduce the E2E \emph{latency}  with improved throughput performance. SDN and NFV based 5G architecture with enhanced  programmability of the network fabric, decoupled network functionalities from hardware, separated  control plane from data plane, and centralized  network intelligence in the network controller is presented in \cite{SDNFV}. In \cite{NFV}, an  information centric scheme is presented in order to integrate the wireless network virtualization with information centric network (ICN). \textcolor{black}{In this architecture, key components such as wireless network infrastructure, radio spectrum resource, virtual resources (including content-level slicing, network-level slicing, and flow-level slicing), and information centric wireless virtualization controller have been  introduced which can support low latency services.}


\textcolor{black}{An NFV-based EPC is introduced in  \cite{taleb2015ease} which is an EPC as a service to ease mobile
	core network (EASE). In this scheme, the elements of the EPC are visualized using VMs. In  \cite{jain2016comparison}, a simple implementation for EPCaaS is proposed in which one of the drawbacks is the increment in the latency between the EPC and virtual network function components. To address such an issue, in \cite{6963800}, the main idea is to partition the virtual network functions into several subsets/groups based on their interaction and workload in order to reduce the network latency. It should be noted that  employing the decentralized control plane and having it closer to the users at the network edge can be helpful for applications with high mobility and low latency. However, it can introduce some issues related to policies and charging enforcements which should be considered for network design based on the requirements. In \cite{LatencyawareNFV}, authors presented  the optimization problem of composing, computing and networking virtual functions to select those nodes along the path that minimizes the overall \emph{latency}  (i.e., network and processing latency). The optimization problem is formulated as a resource constrained shortest path problem on an auxiliary layered graph followed by initial evaluation. }

\begin{table}[!bp]
	\centering
	\caption{\textsc{\textcolor{black}{Proposed Core Network Entities for 5G Visions.}}}
	\label{corenetworkentity}	
	\begin{tabular}{p{1.6 cm}|p{0.8cm}|p{0.8cm}|p{0.8cm}|p{1 cm}|p{0.8 cm}}
		\hline \textbf{Reference} & \textbf {SDN} & \textbf {NFV} & \textbf {MEC} &\textbf {Fog networks}&\textbf {SON}\\ \hline
		\hline \cite{7010528,marquezan2016understanding},\cite{moradi2014softmow, moradi2014softmow2,thyagaturu2016sdn} & \CheckmarkBold & & & &\\
		\hline  \cite{EnablingLowLatency}  & \CheckmarkBold  & & \CheckmarkBold & \CheckmarkBold &\\
		\hline  \cite{SDNAr, SDN5G,SDNtactile, heinonen2014dynamic} & \CheckmarkBold  & &  & &\\ 
		\hline \cite{SDNFV},\cite{costa20175g, wang2017efficient,jain2016comparison} & \CheckmarkBold  & \CheckmarkBold & &  &\\
		\hline \cite{ford2017provisioning} &\CheckmarkBold &\CheckmarkBold&\CheckmarkBold & &\\
		\hline \cite{NFV,taleb2014toward}   && \CheckmarkBold & &\CheckmarkBold & \\
		\hline \cite{cau2016efficient,6963800}   && \CheckmarkBold & & & \\
		\hline \cite{LatencyawareNFV}   & & \CheckmarkBold &\CheckmarkBold &  &\\ 
		\hline \cite{mobiltymnage}   & &\CheckmarkBold &\CheckmarkBold   & &\\ 
		\hline \cite{imran2014challenges} & & & &  & \CheckmarkBold\\ \hline
	\end{tabular} 
\end{table} 

\begin{table*}[!bp]
	\centering
	\caption{\textsc{Overview of Literature in Backhaul Solutions to Achieve Low Latency.}}
	\label{Table_backhauling}
	\begin{tabular}{p{1.2cm}|p{1.6cm}|p{13.5cm}}
		\hline \textbf{Category} & \textbf{Reference} & \textbf{Approach/Techniques}  \\ \hline \hline
		General backhaul & \cite{heinonen2014dynamic} & A dynamic GTP termination scheme combining cloud based GTP with a quick GTP tunnel with a dedicated hardware.\\   
		\cline{2-3}  & \cite{EnablingLowLatency} & GTP tunnel optimization by a new component in 5G complaint network consists of fog networks, MEC and SDN. \\ 
		\cline{2-3} & \cite{backhaulfog} & Modified VLC technology to set up an OW link for low-cost back hauling of small cells. \\ 
		\cline{2-3} & \cite{opticalbacjhaul1} & PON-based architecture with a tailored dynamic bandwidth allocation algorithm. \\ 
		\cline{2-3} &  \textcolor{black}{\cite{gonzalez20165g}} &  \textcolor{black}{MAC-in-MAC Ethernet based unified packet-based transport network.}\\
		\cline{2-3} &  \textcolor{black}{\cite{fiorani2015design}} &  \textcolor{black}{The first architecture is based on over provision transport network whereas the second one is based on dynamic sharing, SDN and NFV controllers.}\\ 
		\cline{2-3} & \cite{7514219} & SDN and cache enable architecture for limited backhaul secererio.\\
		\hline
		mmWave backhaul &	 \cite{weiler2014enabling} & mmWave for fronthaul and backhaul, and split of control and user plane.\\
		\cline{2-3} & \cite{gao2015mmwave} & A digitally-controlled phase-shifter network based hybrid precoding/combining scheme for mmWave massive MIMO.\\
		\cline{2-3} & \cite{taori2015point} & A framework supporting of in-band, point-to-multipoint, non-Line-of-sight and mmWave backhaul.\\
		\cline{2-3} & \cite{levanen2014radio} & A mmWave based backhaul frame structure in 3 - 10 GHz carrier frequencies.\\ 
		\cline{2-3} &  \textcolor{black}{\cite{7503119, 8025130}} & \textcolor{black}{Ultra dense wavelength
		division multiplexing (UDWDM) passive optical networks (PONs) based backhaul solution for mmWave networks.}\\
		\hline
	\end{tabular}
\end{table*}

\textcolor{black}{ In \cite{imran2014challenges}, the authors proposed a detailed approach to implement  the big data empowered self organizing network (SON) 
	in 5G. The procedure to employ the data based on machine learning and data analytics is demonstrated in order to
	create E2E visibility of the network for implementation of a more efficient SON. This approach can meet the stringent 5G requirements
	such as low latency.} \textcolor{black}{In \cite{thyagaturu2016sdn}, the smart gateway (Sm-GW) is employed for scheduling the uplink transmissions of
	the eNBs. Based on simulations, it is demonstrated that the Sm-GW
	scheduling can allocate the data rate  in uplink transmission to the
	eNBs in a fair manner along with reducing packet delays. Traffic of heavily loaded eNBs can make the buffer of the queue of a Sm-GW full which results extra latency. This situation happens due to the massive number of connected eNBs to a single Sm-GW. However, by using effective scheduling, connection with Sm-GW can be distributed among eNBs while maintaining QoS.}
The proposed core network entities for 5G vision are summarized in Table \ref{corenetworkentity}.

\subsection{Backhaul  Solutions}

Backhaul  between base stations and the core network carries the signaling and data from the core and the Internet. Due to the enormous number of small cells and macro cell base stations supporting 1000x capacity, massive connectivity and latency critical services in 5G, the capacity of backhaul is a bottleneck for achieving low \emph{latency}. At current scenario, microwave, copper and optical fiber links are used for backhaul connections based on availability and requirements. 5G backhaul requires higher capacity, lower latency, synchronization, security, and resiliency \cite{backhaulsurbey}. Referring to Table \ref{Table_backhauling}, we divide existing backhaul solutions into 2 parts: (1) General backhaul and (2) mmWave backhaul. The solutions are described as follows.

\subsubsection{General backhaul} General backhaul includes a dynamic GPRS tunneling protocol (GTP) termination mechanism that combines cloud based GTP with a quick GTP tunnel  proposed in \cite{heinonen2014dynamic}. Based on  the user request or other factors, the system can change its mode from a cloud-based GTP tunnel to the quick GTP tunnel. In \cite{EnablingLowLatency}, a 5G vision compliant architecture is presented to reduce \emph{latency}  combining with the fog network, MEC and SDN. The optimization of  GTP tunnels is accomplished  by a novel element acting as an intermediate node between eNB and the mobile network interface accompanied with the Internet. In \cite{backhaulfog}, modified VLC technology is used to set up an optical window (OW) link for low-cost backhaul of small cells to achieve a \emph{latency}  of 10 ms. Moreover, using a next generation baseband chipset, E2E latency below 2 ms can be achieved. 
An efficient PON-based architecture is proposed in \cite{opticalbacjhaul1} that offers ultra-short \emph{latency}  for handovers
by enhancing connectivity between neighboring cells. Additionally, the authors propose  a tailored dynamic bandwidth allocation
algorithm for a fast handover between eNBs, which are associated to the same or diverse optical network units.
\begin{table*}[!bp]
	\centering
	\caption{\textsc{Overview of Literature in Caching.}}
	\label{Table_caching_overview}
	\begin{tabular}{p{2.9 cm}|p{3cm}|p{11.2 cm}}
		\hline \textbf{Aspect}  &  \textbf{Reference} & \textbf{Summary} \\ \hline \hline
		Content caching & \cite{22c,24c,26c,27c,28c,29c,31c,32c,33c,35c,36c,37c,38c,39c,40c,41c,42c,43c,44c,45c} & Filling of  appropriate data is investigated by diverse techniques employing time intervals in which the network is not congested.\\ \hline
		Content delivery & \cite{23c,24c,25c,28c,29c,30c,31c,32c,33c,35c,36c,37c,38c,39c,40c,41c,42c,43c,44c,45c}& Content delivery to requested users is presented by different approaches for reduction of latency.\\ \hline
		Centralized caching& \cite{22c,23c,24c,25c,26c,29c,32c,35c,36c,37c,38c,39c,40c,41c,43c,44c,45c}& Various centralized caching is investigated with assumption that a coordinator with access to almost all the information about the storage capacities of different BSs, the connectivity of the users and BSs, and etc.\\ \hline
		Distributed caching& \cite{27c,28c,30c,31c,33c,42c,45c}&  \textcolor{black}{Various aspects of distributed caching has been investigated in order to minimize the communication overhead among SBSs and the central scheduler.}
		
		\\ \hline
		
		Latency-Storage trade-off & \cite{35c,36c,37c,38c,39c,40c, girgis2017decentralized, sengupta2016cloud} & Fundamental trade off between storage and latency is investigated in radio networks complemented with  cache-enabled nodes. \\ \hline
		
	\end{tabular}
\end{table*}

\textcolor{black}{ In transport networks, latency requirement plays a key role. For instance, the main requirement for  machine type communications is the latency that should be kept as low as possible. Therefore, efficient design of  5G transport
	networks is critical.} \textcolor{black}{In \cite{gonzalez20165g}, a detailed perspective  of the 5G crosshaul design is proposed in order to  introduce the key goal  of
transporting the backhaul and fronthaul traffic in a unified packet-based transport network based on
MAC-in-MAC Ethernet. Moreover,  the SDN/NFV-based 5G-crosshaul control plane architecture is investigated which decouples the logically centralized control plane and data plane. This can contribute in latency reduction aspect. In \cite{fiorani2015design}, two candidate technologies for 5G transport networks are presented. One of them is based on the over-provisioning of transport resources while the second
architecture is based on dynamic resource sharing and NFV/SDN-based controller to handle the latency requirements.} \\

\textcolor{black}{In \cite{7514219}, SDN and cache enabled heterogeneous network is proposed where C-plane and U-plane are split. The caches of macro  and small cells are overlayed and cooperated in a limited backhaul scenario while ensuring seamless user experiences for coverage, low latency, energy efficiency and throughput. In \cite{fiorani2015design}, two candidate technologies for 5G
transport networks is presented to handle the latency requirements in which  one of them is based on the
over-provisioning of transport resources, while the second
architecture is based on dynamic resource sharing and
NFV and SDN-based controller. In \cite{7514219}, SDN and cache enabled erogenous network is proposed where C-plane and U-plane are spitted. The caches of macro  and small cells are overlayed and cooperated in a limited backhaul scenario while ensuring seam user experiences such as coverage, energy efficiency and throughput.}\\

\begin{figure*}[!tp]
	\centering 
	\includegraphics[width=1\linewidth]{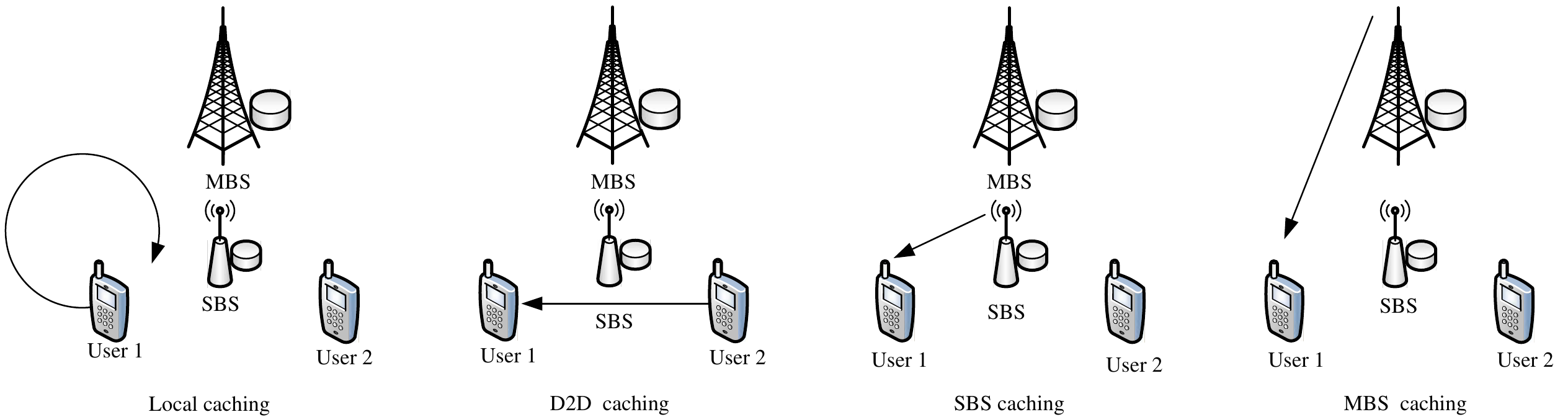}
	\caption{Different types of caching in 5G.}
	\label{cachingtypes}
\end{figure*}

{\color{black} \subsubsection{mmWave Backhaul} In addition to the presented solutions for backhaul, mmWave employment in backhaul can be considered as a promising solution for \emph{latency}  reduction. In order to have the 
	enhanced user experience, the BSs should  be in touch with core network and all other BSs via a 
		low \emph{latency}  backhaul \cite{dehos2014millimeter}.  In \cite{weiler2014enabling},  the authors proposed a scheme that employed mmWave links as backhaul,
		fronthaul and access in which  a new separation method between control and user plane  is proposed  for 5G cellular network. A reasonable split among control and user plane improve the user QoS by providing ubiquitous  high data
		rates in mmWave SBS coverage.

In \cite{gao2015mmwave}, to implement  an ultra-dense network (UDN) for the  future 5G network and  providing high data rates, the need of a reliable, gigahertz bandwidth,
		and economical backhaul is emphasized. Since mmWave can be easily integrated with
		massive MIMO to improve link reliability, and  can provide sufficient  data rate for wireless
		backhaul, it is a promising candidate for such a scenario. Considering a massive MIMO scenario, a hybrid precoding
		approach is  considered, in which each BS can  cover
		several SBSs with multiple streams for each SBS at the same time. In \cite{taori2015point}, the authors proposed a
	solution framework for
	supporting an in-band, point to multi-point, NLOS,
	mmWave backhaul in order to provide a cost-effective and low \emph{latency}  solution for wireless
	backhaul. It is shown that an in-band wireless
	backhaul for inter BS coordination is
	feasible while the cell access capacities are not affected considerably. 
	
	In \cite{levanen2014radio}, a 
		frame design for  mmWave  communications is proposed  for 5G SBS
		network radio interface in 3-10 GHz.
		For both of LOS and NLOS scenarios different frame designs are proposed, which have a  frame duration of 0.1 ms and  0.05 ms, respectively, to achieve low \emph{latency}. The proposed LOS structure can be  assumed as a suitable solution  for
		short distance indoor wireless access or in-band backhaul.}
\textcolor{black}{In order to obtain high capacity and low latency backhauling, the EU research project 5G STEP-FWD introduced a novel
design that deploys ultra  dense wavelength
division multiplexing (UDWDM) passive optical networks (PONs) as the backhaul of mmWave networks. The proposed {\color{black}scheme is }based on  the ultra-narrow
wavelength spacing of the UDWDM technology to provide seamless connectivity {\color{black}for  dense} small-cell
networks~{\color{black}\cite{thyagaturu2016software, 8025130}.}}

\section{Caching Solutions for Low Latency}
In addition to the shortage of the radio spectrum, the insufficient  capacity of backhaul links can be considered as a bottleneck for low latency communication. The long delay can be due to the requests of too many users in peak-traffic hours. Thus, latency reduction is crucial for users’ QoS and QoE in the 5G networks. Caching and in a more general category, information centric networking, can be assumed as one of the promising candidate technologies to design a paradigm shift for latency reduction in next generation communication systems \cite{cachingsurvey1,cachingsurvey2}.

In this section, referring to Table \ref{Table_caching_overview}, we present a detailed overview of caching concepts for cellular network followed by fundamental limits and existing solutions.

\subsection{Caching for cellular network}

Let us consider, a scenario that a user requests content from a content library  \textcolor{black}{$F=\{f_1,f_2,......,f_k\}$,} where $k$ is the number of files. The files are sorted with popularity where $f_1$ and $f_k$  are the most and least popular files, respectively. The popularity of a requested file $l$ can be written  \cite {chen2016mobility} as

\textcolor{black}{
	\begin{equation}
	\phi_l=\frac{l^{-\gamma}}{\sum_{i=1}^{k} l^{-\gamma}},
	\end{equation}}

\noindent where $l \in \{1, 2, ..., k\}$ and \textcolor{black}{ $\gamma$ is the parameter for uneven distribution of popularity in $F$ which follows Zipf distribution.}
For $N$  eNBs $\mathcal{B}=\mathrm{\{BS_1,BS_2,.......,BS_N\}}$ with each eNB having capacity ${C}$, the probability of caching of  file $f_l$ by an eNB can be obtained as

\textcolor{black}{
	\begin{equation}
	P_l^\mathcal{B}=1-e^{-\rho \sigma \pi R^2},
	\end{equation}}
	
\noindent where $\rho$ is the spatial density  of eNBs following a Poisson point process \cite{CoperaCaching, Spatialspectrum, 6190344}, and
$\sigma$ is the probability that   file $f_l$ is cached  within $\mathcal{B}$.
Then, the total probability of getting content from the eNB can be written as 

\begin{equation}\label{99}
P^\mathcal{B}=\sum_{i=1}^{N} \phi_i P_i^\mathcal{B}.
\end{equation} 

\textcolor{black}{The probability of getting the content  as in \eqref{99} is directly associated with the latency of downloading it, and hence, effective caching strategies can help in significantly reducing latency in 5G networks}

\begin{table}[!tp]
	\centering
	\caption{\textsc{Different Types of Caching Schemes for the Cellular Network.}}
	\label{Cachingdefinition}
	\begin{tabular}{p{1.3 cm}|p{1.2cm}|p{5.2 cm}}
		\hline
		\textbf{Cases}  & 	\textbf{Reference}  &\textbf{Description}    \\ \hline \hline
		\vskip 6pt Local caching  &\vskip 6pt \cite{6736746} & When a UE wants to access a content, it first checks in itself. Once the content is confirmed in the local caching storage, it is accessed by the UE without any delay.
		\\ \hline
		\vskip 12pt D2D caching & \vskip 14pt \cite {chen2016mobility} &  If the requested content is not found locally, user will seek it within the range of it's D2D communication. If it is found in nearby devices, it is delivered to the requester UE by D2D communication.
		\\ \hline
		\vskip 0pt  SBS caching & \vskip 4pt \cite{SBScaching}& If the requested content is available in the local SBS, the content is delivered to the UE by the local SBS.
		 \\ \hline
		\vskip 3pt	MBS caching &\vskip 3pt  \cite {chen2016mobility} &If the content is not found in local caching storage, nearby devices or SBS caching, the content is delivered by MBS caching.
		\\ \hline	
	\end{tabular}		
\end{table}

The proposed caching schemes for mobile networks can be divided into 4 categories: (1) Local caching, (2) Device to device (D2D) caching, (3) SBS caching, and (4) Macro base station (MBS) caching. Each of these caching types can reduce the \emph{latency}  by providing the requested content for the users using a way other than bringing  it from the core network using backhaul links. In fact, each user starts from the nearest  source to look for its desiered content and proceed until finding it in any of the proposed sources. The different types of caching for cellular network are illustrated in Fig. \ref{cachingtypes} followed by the summarized descriptions presented in Table \ref{Cachingdefinition}.

\subsection{Fundamental Latency-storage trade-off in Caching}

There are several fundamental limits for caching in mobile networks including latency versus storage, memory versus rate \cite {41c}, memory versus CSIT \cite {42c}, storage versus maximum link load \cite {43c}, and caching capacity versus delivery rate \cite {44c}.  
As defined in Table~\ref{Metricsdefinition}, from an information theoretic point of view, authors employed the metrics such as normalized delivery time (NDT), fractional delivery time (FDT), and delivery time per bit (DTB) for investigation of the latency storage trade-off in caching networks. In most of these works \cite{27c,28c,30c,31c,33c,42c,45c}, for a given scenario, an upper bound or lower bound for the considered metric is derived in order to get useful insights of this trade-off. The summary of \emph{latency}  storage trade-off works is presented in Table \ref{latencystorage}.

\begin{table}[!bp]
	\centering
	\caption{\textsc{Definition of the Metrics used for Latency Evaluation in Caching Schemes.}}
	\label{Metricsdefinition}
	\begin{tabular}{p{0.7 cm}|p{1.2 cm}|p{5.7 cm}}
		\hline
		\textbf{Cases}  & \textbf{Reference} & \textbf{Defenition}    \\ \hline \hline
		\vskip 2pt NDT& \vskip 2pt \cite{35c}  & Defined as asymptotic delivery delay per bit in the high-power, long-blocklength case.
		\\ \hline
		\vskip 5pt DTB &\vskip 5pt \cite{40c}&  Defined as the ratio between  the duration of
		transmission in channel to the file size in bits for the very large file size regime.
		\\ \hline
		\vskip 4pt  FDT &   \vskip 4pt \cite{39c}& Defined as the \textcolor{black}{  worst-case delivery latency
			for  the real  load at a rate described by the 
			DoF of the  channel. }\\ \hline
		
	\end{tabular}		
\end{table}
\begin{table}[!tp]
	\centering
	\caption{\textsc{Summary of the works on latency-storage trade-off in caching.} }
	\label{latencystorage}
	\begin{tabular}{p{0.7 cm}|p{0.8 cm}|p{6 cm}}
		\hline
		\textbf{Ref.}  &\textbf{Used metric} & \textbf{Description}    \\ \hline \hline
		\cite{35c}  &NDT& Lower bound for NDT is derived for a   general cache-enabled  network for both perfect and imperfect CSI.
		\\ \hline
		\cite{36c} &NDT&  The trade-off between NDT and front-haul
		and caching resources is characterized and optimal caching front-haul transmission is obtained.
		\\ \hline
		\cite{37c} &FDT& \textcolor{black}{ For a  $3 \times 3$ wireless interference network the storage-latency trade-off is investigated
			while all transmitters and receivers are equipped with caches.}
		\\ \hline
		
		\cite{39c} &FDT&  For a scenario with a $3 \times 3$ MIMO system in which the nodes are enabled with several antennas, the trade off between storage and latency is investigated.
		\\ \hline	
		\cite{40c} &DTB&  DTB is used to characterize the system
		performance as a \textcolor{black}{ function of cache storage and capacity of
			backhaul links connected to SBS. }
		\\ \hline	
		\cite{girgis2017decentralized} &NDT&  The trade-off
		between storage and latency for a distributed caching scenario in
		fog radio access networks is characterized.
		\\ \hline	
		\cite{sengupta2016cloud} &NDT&  Cloud-based compressed precoding and edge-based interference management introduced as two
		major techniques for \textcolor{black}{ optimal performance of cloud and caching resources
			in different cases. }
		\\ \hline	
		\cite{goseling2017delivery} &NDT&  Considering heterogeneous timeliness requests depending on
		application, the fundamental trade-off between the
		delivery latencies of different users’ requests is characterized
		using NDT. 
		\\ \hline	
		\cite{koh2017cloud} &NDT&  Upper and lower bounds for minimum delivery latency as a function of cache and
		fronthaul resources is obtained  over fronthaul and wireless link in a F-RAN
		with a wireless multicast fronthaul. 
		\\ \hline
		
		\cite{7849002} &NDT& Assuming an F-RAN when pipelined fronthaul edge transmission is used, lower and upper bounds on the NDT is presented.
		
		\\ \hline
		\cite{azimi2017online} &NDT&  Assuming both content placement and delivery phases in one time slot, it is shown that the proposed approach outperforms  the offline
			caching.
		\\ \hline	
		
	\end{tabular}		
\end{table}


The authors in \cite{35c} investigated the storage latency  trade-off using a new metric called NDT. This metric measures the worst-case \emph{latency}  that can happen in a cache-aided wireless network divided by that of an ideal system with unlimited caching capability. Considering a general cache-aided wireless network, the lower bound for NDT is presented in terms of the ratio of the existing file memory at the edge node and the total size of files for both perfect channel state information (CSI) and imperfect  CSI. 

Authors in \cite{36c} employed NDT as well in order to characterize the trade-off between NDT and fronthaul/caching resources. Using this information-theoretic analysis of fog radio access networks, optimal caching front-haul transmission is obtained. 
In \cite{37c}, the \emph{latency}  storage  trade-off in a $3\times3$ wireless interference network is investigated while all transmitters and receivers are equipped with caches. Another metric called (FDT) is proposed in order to characterize the trade-off between latency and storage. This information theoretic performance metric is actually a refined version of the metric originally proposed in \cite{35c}. The FDT can reflect the load reduction as well. In a similar work \cite{38c} the well-known DoF metric is used which does not reflect the load reduction. Moreover, the proposed approach in \cite{38c}
just considers the one-shot linear processing, but  interference alignment scheme in  \cite{35c}  may require infinite symbol
	extension. 

In \cite{38c}, an optimization problem is designed to minimize the number of required communication blocks for content delivery. Then,  a lower bound is proposed on the value of the objective function. Using the same metric, the authors in \cite{39c} investigated the fundamental trade-off for a cache-enabled MIMO system. Considering a scenario with a $3 \times 3$ MIMO system in which the nodes are enabled with several antennas, the trade off between storage and latency is investigated. In addition to FDT and showing its optimality   for some ranges of cache size,  the model can consider the effect of real traffic load at a rate specified by the DoF of the  channel.

In \cite{40c}, a cellular network is considered with multiple SBSs with limited cache capacity  in which there is interference among them. Here, another information theoretic metric based on delivery latency is defined as well in order to characterize the system performance as a function of SBS cache memory and capacity of backhaul links connected to SBS. Using this metric which is called DTB, the trade off between latency and system resources is investigated.  In \cite{girgis2017decentralized}, using NDT trade-off between storage and latency a distributed
caching scenario in fog radio access networks is characterized. In the presented approach, a coded delivery scheme is proposed to minimize the \emph{latency}  for delivering user
demands for two edge-nodes
and arbitrary number of users. It is shown  that    using decentralized
	placement, the presented delivery approach can obtain a considerable
	performance improvement in comparison to  the derived lower bound. 

In \cite{sengupta2016cloud}, again NDT is employed to characterize the fundamental trade off between delivery
	latency and  system architecture.
	Considering NDT as the criterion for latency evaluation of the system, some bounds
	on its value are proposed. In the light of such bounds, useful insights on
	the latency and storage trade off are obtained.
	It is demonstrated that  in order to obtain the lowest delivery \emph{latency}, cloud-based compressed precoding and edge-based interference management should be considered as two major techniques for  optimal performance of cloud and caching resources
	in different cases. In \cite{goseling2017delivery},  for a fog radio access network, the heterogeneous timeliness requests depending on application is considered while in the existing works
the assumption is that all requests
have identical \emph{latency}  for all files in the content library. The fundamental trade-off between the delivery latencies of different
users' requests is characterized using NDT. The minimization of the
average delivery \emph{latency}  as a function of the content popularity
profile remains as a future research direction.

In \cite{koh2017cloud}, the total delivery latency over
fronthaul and wireless link in a fog radio access network with a wireless multicast fronthaul is investigated. Again using NDT, the  optimal delivery
	\emph{latency}  based on  cache storage and fronthaul resources is formulated and
upper and lower bounds are obtained. It is shown that in contrast to  the receiver-side caching, coded multi-casting can not help in
decreasing the NDT when
two users and two edge nodes (ENs) are available.  In \cite{7849002}, for a F-RAN, the NDT is used to characterize the   performance of a fog radio access network when pipelined fronthaul edge
transmission is used.

In \cite{azimi2017online}, a F-RAN system is considered in which ENs
are cache-enabled with limited storage. On the contrary to the existing works that focused on offline caching where both caching phases are considered separately, the proposed method can integrate both of them in each time slot. 
The performance is characterized using NDT and compared to that of optimal offline caching schemes. It is shown that the proposed approach outperforms the offline caching.

\subsection{Existing Caching Solutions for 5G} In general, the file delivery service
in mobile networks can be classified into two parts: {\em cache
	placement}, and {\em content delivery} \cite{fang2014survey}.
In cache placement, the cached content on the BSs is determined, which is usually based on the amounts of requests from users. Cache placement
can be done using a centralized or distributed approach. In centralized approach, a coordinator is assumed  with access to almost all the information about the \textcolor{black}{memory size of  BSs, the connectivity}  of
the users, and the BSs. However, in some scenarios that there is no central controller, these schemes can not be applicable and a distributed cache placement is required \cite{27c}. In each of the caching phases whether centralized or distributed, the design can reduce the delivery time of the requested content and latency of the system.

There are some efforts in the literature on investigation of different challenges raised up in centralized cache placement problems.
In \cite{22c}, the cache placement problem was investigated in a scenario including SBSs, called helpers with weak backhaul links but large memory size. In experimental evaluation, it is shown that the proposed scheme can achieve a considerable performance improvement for the users at reasonable QoS levels. In \cite{26c}, authors aim at minimizing the average download delay of wireless caching networks with respect to caching placement matrix. It is demonstrated  that the backhaul propagation delay can affect the caching placement. 


In some scenarios, it is more desirable to  design the caching problems in a distributed approach. In \cite{27c}, a distributed cache placement approach is proposed in order to minimize the average download delay while some constraint for BSs storage capacities are met. The formulated optimization problem which is NP-hard is solved using a belief propagation based distributed algorithm with low complexity. This optimization problem makes sense because there is a trade-off between latency and storage capacity in caching networks. The comparison between the performance of the proposed distributed scheme with that of the centralized algorithm in \cite{22c} is presented as well. In \cite{28c}, a decentralized content placement caching scheme is presented. Although there is no coordination, the proposed scheme can \textcolor{black}{attain a rate as good as the optimal centralized scheme proposed in \cite{22c}}. In \cite{34c}, two caching and  delivery schemes are considered. The first one operates in a centralized manner, while the second one is based on decentralized caching. For both cases, the trade-off between coded multi-casting and spatial reuse is reflected by the code length.

\begin{table*}[!hbp]
	\centering
	\caption{ \textsc{Platform and Performance on Field Tests, Trials and Experiments.}}
	\label{Filedtests}	
	\begin{tabular}{p{1.1 cm}|p{1.8cm}|p{0.5cm}|p{0.5cm}|p{1cm}|p{4cm}|p{6cm}}
		\hline \textbf{Reference} & \textbf {Evaluation methodology} & \textbf {SDR} & \textbf {DSP} & \textbf {mmWave} & \textbf {Conventional/ Proprietary LTE } &\textbf {Remarks} \\ \hline \hline
		\cite{SDRAplatform} & &  \CheckmarkBold & & & & \textcolor{black}{Trials of 5G concepts along with a novel air interface} \\ \hline
		\cite{Methology} & \CheckmarkBold & & & & & 4 test cases and 15 KPIs is proposed.\\ \hline
		\cite{5Gtrailthrouput} & &   & & \CheckmarkBold & & mmWave aggregation\\ \hline
		\cite{5Gradio} & &   &\CheckmarkBold &  & & \textcolor{black}{RTT latency of 1 or 2~ms is achieved. Moreover, to achieve latency on the order of couple of hundreds microseconds  over air interface, cross later approach is recommended.}\\ \hline
		\cite{testbedv2x}  & & \CheckmarkBold   & &  & & Low latency VANET\\ \hline
		\cite{fiberoptics}  & &   & \CheckmarkBold &  & & \textcolor{black}{DSP round-trip latency less than $2\mu$s is achieved for channel aggregation and de-aggregation for 48 20~MHz LTE signals. }\\ \hline
		\cite{Improvinglatencyreliability}  & & \CheckmarkBold  &  &  & & \textcolor{black}{20 times latency reduction in comparison to existing works} \\ \hline
		\cite{7794604} & &  &   & \CheckmarkBold & & Minimum latency 3 ms \\ \hline
		\cite{5Gradio} & &  & \CheckmarkBold &  & & Latency $\leq$ 1 or 2 ms\\ \hline
		\cite{7386169} & &  &  &  & \CheckmarkBold & \\ \hline
		\cite{7833471} & &  &  &  & \CheckmarkBold & Latency $\leq$ 17 ms \\ \hline	
		\cite{GuanZRTBSK16} &\CheckmarkBold &  &  &  &  &  HARQ RTT $\le$ 1.5 ms \\ \hline
		\cite{kela2015novel} & \CheckmarkBold&  &  &  &  & RTT latency $\leq$ 1 ms \\ \hline
		
	\end{tabular} 
\end{table*}

In addition to the aforementioned literature, in \cite{jiang2017optimal}, content caching, and content delivery  schemes are proposed considering cooperation to address the explosive enhancements of demand for mobile network applications. Defining the objective as minimizing the average downloading \emph{latency}, it is demonstrated that the proposed content assignment and delivery policy scheme has a better performance in comparison to the  previous known content caching schemes in terms of average downloading \emph{latency}. 
A weighted optimization problem is formulated in \cite{29c} to minimize the traffic of backhaul and downlink while the constraints for cache memory size and bandwidth limitation for D2D communication is considered. It is shown in \cite{30c} that if latency awareness is considered in caching management, it is an effective approach to reduce the delivery time of \emph{latency}  sensitive applications, and the global delivery time of users. In the proposed model, two main advantages is claimed. First, it not only has a better performance in term of delivery time at the end-user, but also affects the link load reduction. Second, a faster convergence with respect to probabilistic caching is achieved. In \cite{31c}, the effect of joint latency awareness and forwarding is investigated in a cache-enabled network. Authors proposed a scheme which is based on caching and forwarding strategies in order to  improve E2E experienced latency  by the UEs while there is no coordination among them. 

In \cite{32c}, a cooperative content caching approach between BSs in cache-enabled multi-cell network is considered. Due to the trade-off between storage and latency, cooperative caching optimization problem is designed in order to minimize the average delay while a constraint on the finite cache size at BSs is met. It is shown that cooperation among cells can considerably reduce delay in comparison to that of non-cooperative case. Moreover, the gains of the proposed scheme will be increased in more diverse and heavier load traffics. In \cite{33c}, the aim of the work is to minimize the data transmission delay for the P2P caching system while considering the effect of cache size, all mobiles in a cell are considered as several P2P caching groups.  Then, the problem is formulated as a stochastic optimization problem and solved using Markov decision process (MDP) to obtain the optimal solution. 

In \cite{7881651}, a cooperative multicast-aware caching strategy is proposed for the BSs to decrease the
	average latency of content delivery  in 5G cellular networks. The proposed scheme is carefully designed in order to take into account the benefit of multicast and cooperation while in the existing caching schemes, the popular content simply is brought close  to the users. The optimization problem is  formulated in order to minimize average latency for all the content requests. It is demonstrated that via various trace-driven simulations that the proposed cooperative  multicast-aware caching scheme can provide up to 13\% decrease in the average content-access latency in comparison to  multicast-aware caching scheme with the same total cache capacity.

In \cite{7247328}, the authors presented a cooperative caching
architecture in which multiple locally cache-enabled nodes of  cloudlets
	interact cooperatively in a decentralized cloud service networks. By proposing a
	content distribution strategy, the problem is formulated so that  the mean total content
	delivery delay for all users in the proposed scheme is minimized. It is confirmed that the approach can enhance the cache hit rate, and also reduce the the content
delivery latency  in comparison to  existing solutions. In \cite{7752660},   the E2E  packet transmission in a cache-enabled network is modeled in which both the wired backhaul and the 
RAN are jointly considered. The performances
of both the on-peak  and the off-peak network are
investigated while  both   the wired
backhaul and the RAN are considered. The E2E average packet latency is elaborated  with the change of the request rate.
It is shown that the average packet \emph{latency}  reduces  in comparison 
to that of the system without caching ability due to the
traffic offloading of the wired backhaul via caching.

\section{Field Tests, Trials and Experiments}

In this section, we present some representative field tests, trials and experiments for 5G low \emph{latency}. The related literature is  summarized in Table \ref{Filedtests}, where each of the individual references  will be described in further detail below.

The study \cite{SDRAplatform} presents SDR based hardware platform to verify the concept of 5G. This facilitates initial proof-of-concepts (PoC) of novel 5G air interface and other concepts by extending hardware-in-the-loop (HIL) experiments to small laboratory experiments and finally trials of outdoor tests. Such an SDR based hardware can demonstrate high-capacity, low \emph{latency} and coverage capabilities of LTE-A solutions. In \cite{Methology}, evaluation methodology including some novel test environments and certain new key performance indicators are discussed in order to evaluate 5G network. Here, four candidate test environments such as indoor isolated environment and high speed train environment, and fifteen key performance indicators such as latency, throughput, network energy efficiency and device connection density are emphasized  for performance evaluation.

In \cite{5Gtrailthrouput}, 5G system operating at 15 GHz is presented followed by some experimental  results. \textcolor{black}{Here 0.2 ms subframe (14 \textcolor{black}{OFDM symbols}) is used for throughput, latency and other performance evaluation.} The hardware implementation results of digital signal processing (DSP) and SDR based 5G system for \emph{low latency} is presented in \cite{5Gradio}. In this study, both the short TTI (sTTI) frame structures and wider subcarrier spacings are implemented in DSP platform. Based on the configurations of the system, RTT latency as low as 1~ms can be achieved. However, for achieving latency on the order of a few  $\mu s$, optimization in between controllers and processing machines needs to be performed by cross-layer fashion. Additionally, the tail latency is argued to be considered in strict latency requirements assessment while maintaining required reliability.

An SDR based test bed is presented in \cite{testbedv2x} for cooperative automated driving with some experimental results from lab measurements. It implements flexible air interface consisting of re-configurable frame structure with fast-feedback, new pulse shaped OFDM (P-OFDM) waveform, low \emph{latency} multiple-access scheme and robust hybrid synchronization, which ensure low latency high reliable communication. Results of the experimental trials are presented in \cite{fiberoptics}, which utilizes DSP techniques for channel aggregation and de-aggregation, adjacent channel leakage ratio reduction, frequency-domain windowing, and synchronous transmission of  I/Q waveforms and code words used in control and management function. In the proposed experiment, transmission of 48 chunks of 20 MHz LTE signals using a common public radio interface of capacity 59 Gb/s can achieve RTT DSP latency of less than $2~\mu$s and  mean error-vector magnitude of about $2.5\%$ after fronthaul fiber communication. This mobile fronthaul technique shows the path towards ultra low latency  integrated fiber/wireless access networks.

In  \cite{Improvinglatencyreliability}, a multi-terminal massive SM-MIMO system is evaluated considering realistic scenarios. The authors developed a massive SM-MIMO OFDM system prototype utilizing  multiple off-the-shelf SDR modules which serve as IoT terminals. Two linear detection schemes with diverse complexity levels were tested for detection instead of maximum likelihood detection (MLD) schemes. It demonstrates the similar real-time SINR performance of the MLD techniques along with 20 times \emph{latency} reduction over existing works. The promising results urge the utilization of massive SM-MIMO systems for latency reduction and reliability enhancement in IoT transmissions. In \cite{7794604}, the performance of a lower latency frame structure was evaluated in field tests using a 5G mmWave proof-of-concept (PoC) system. It was found that the slot interleaving frame structure can achieve RTT \emph{latency} of 3 ms in the $70-80\%$ of the trial course. Additionally, beam tracking  and throughput performance were evaluated  in field tests at a speed up to  20 km/h on LOS outdoor environment. It was confirmed that  mmWave system can obtain throughput of 1~Gbps in the $38\%$ of the trial time at 20 km/h speed.
 
 In \cite{7386169}, a low complexity receiver design was introduced followed by verification of superiority of an SCMA system via simulations and real-time prototyping. It can provide up to $300\%$ overloading that triples the whole system throughput while still enjoying the link performance close to orthogonal transmissions. In \cite{7833471}, the concept of MEC was introduced first time for 5G followed by promising field tests. The MEC was tested and analyzed on various cases including  local breakout and network E2E \emph{latency}. It was concluded that MEC can support low \emph{latency} services of not lower than 17 ms. It also urged that stricter requirement of latency needs to be investigated from the new radio technologies or D2D communication. In  \cite{GuanZRTBSK16},  a lab trial is presented to study the feasibility of ultra-low latency for 5G. It is shown that  1.5 ms HARQ
 RTT for TDD downlink in a lab trial is achievable when using the available test equipment in the literature while a novel frame structure and the
associated signaling procedure is employed. The proposed scheme has  5 times better \emph{latency}
performance in comparison to  the existing
LTE-Advanced standard.

In \cite{kela2015novel}, a novel  frame structure  is tested using a proprietary  quasi-static system simulator for ultra-dense 5G outdoor RANs. In this regard, a  frame structure is designed in order to facilitate low latency and multiuser spatial multiplexing on radio interface along with small-scale packet transmission and mobility support. It is found that performance of the introduced 5G network is better than  that of LTE in case of air interface \emph{latency}. In particular, considering  UL scheduling requests in the RTT latency, the proposed frame structure can achieve \emph{latency} as low as 0.8365~ms which is reduced by a factor of 5 in comparison to that of LTE. This satisfies 5G \emph{latency} requirement (i.e.,1 ms latency).


\section{\textcolor{black}{Open Issues, Challenges and Future Research Directions}}
  
  \textcolor{black}{While there are some existing proposals to reduce latency to 1~ms, there are several open issues and challenges for future research. The area of exploration includes RAN, core network, \textcolor{black}{backhauling}, caching and resource management. \textcolor{black}{Also, the existing techniques need to be validated in field tests and should  evolve from current LTE systems}. In the following section, we discuss some of the open issues and challenges \textcolor{black}{which needed to  be explored and addressed by researchers from both academia and industry.}}
  
  \subsection{\textcolor{black}{RAN Issues}}
  
  \textcolor{black}{As discussed in Section~IV, most of the fundamental constraints for achieving low latency requires modification in  PHY and MAC \textcolor{black}{layer which are at} RAN level. Even though several promising solutions are proposed to date, we believe that the following issues at RAN level need to be investigated \textcolor{black}{further}.}
  
  \begin{itemize}
  	
  	\item \textcolor{black}{For achieving low latency in 5G networks, mmWave is \textcolor{black}{a} promising technology which brings massive new spectrum for communications in the 3-300~GHz band. However, mmWave is dependent on diverse aspects such as transmitter/receiver location and environmental topology \cite{6387266,SCMA2}. Moreover, channel modeling with delay spread, path loss, NLOS beam forming, \textcolor{black}{and} angular spread need to be investigated in indoor and outdoor environment which are still evolving~\cite{6515173}. Additionally, more in-depth knowledge of physics behind mmWave regarding \textcolor{black}{aspects} such as Doppler, propagation, atmospheric absorption, reflection, \textcolor{black}{refraction, attenuation, and multi-path should be developed for utilization of mmWave.}}

  	\item  \textcolor{black}{In conventional packet transmission, distortion and thermal noise induced by propagation channel get averaged due to large size of packet \cite{7529226}. \textcolor{black}{However,} in case of small size packet, such averaging is not possible. \textcolor{black}{Thus,} proper channel modeling followed by simulations and \textcolor{black}{field} tests for small packet in diverse carrier bands need be investigated.}
  	
  	
  	\item \textcolor{black}{The challenge of   admission control  in RAN for spectral and energy efficiency with  latency constraint is not well explored~\cite{7925934}.} \textcolor{black}{CRAN/HRAN provides spectral and energy efficiency while aspects such as caching can ensure low latency~\cite{NEMNEM1973}. Researchers can work for performance \textcolor{black}{bounds} regarding this issue.}

  	\item \textcolor{black}{Orthogonality and synchronization is a major drawback of OFDM modulation for achieving low latency. On the \textcolor{black}{other hand}, orthogonality and synchronization are very important for data readability. Recently, different non-orthogonal and asynchronous multiple access schemes such as SCMA, \textcolor{black}{IDMA} and GFDM\textcolor{black}{,} FBMC and UFMC have been proposed. \textcolor{black}{However,} more effective access techniques and waveforms which require less coordination, \textcolor{black}{ensure} robustness in disperse channel, and provide high spectral efficiency is \textcolor{black}{a potential research} area~\cite{7023145,SCMA2}.}
  	
  	\item \textcolor{black}{Low complexity antenna, beam steering large antenna array and efficient symbol detection such as compressed  sensing is conducive for low latency communication. For this, heuristic beam forming design at BS level, beam training protocols, and weight calculation and reliable error correction technique need to be studied~\cite{6736750,SCMA2}. On the other hand, at the receiver level, low complexity sensing \textcolor{black}{techniques} and receiver design should be \textcolor{black}{the} focus of \textcolor{black}{the} future research.}
  	
  	\item \textcolor{black}{One of the major challenges is that latency critical packets are to be multiplexed with other packets. \textcolor{black}{There are} solutions such as instant access for latency critical packets \textcolor{black}{ceasing transmission of} other packets, and reservation of resources for latency critical services~\cite{ji2017introduction}. \textcolor{black}{We believe that these issues are not well-explored and calls for further study.} }
  	
  	
  	\item \textcolor{black}{Even though the \textcolor{black}{main} objective of CRAN is to reduce costs and to enhance energy  and spectral efficiency, it might be combined with heterogeneous networks termed as H-RAN. However, it is  very challenging to design 5G network with large CRAN/HRAN~\cite{NEMNEM1973}. In this case, various \textcolor{black}{trade-offs} \textcolor{black}{including} energy efficiency \textcolor{black}{versus} latency can be investigated.}
  	
 %
  	
  \end{itemize}
  
  \subsection{\textcolor{black}{Core Network Issues}}
  
  \textcolor{black}{In the core network, several new entities \textcolor{black}{such as} SDN and NFV have been introduced for supporting large capacity, massive connectivity and low latency with seamless operation. Since these entities are not part of the legacy  LTE system, extensive works need to be carried out \textcolor{black}{for the} standardization and development in the context of  5G. Some of the issues necessitating research in the core network level are as follows.}
  
  \begin{itemize}
  	\item \textcolor{black}{The main challenge of \textcolor{black}{SDN/NFV-based} core network design is the management and orchestration of these heterogeneous resources~\cite{7494060}. Effective resource allocation and implementation  of functions in this heterogeneous environment while also maintaining low latency is an area of emerging research need.}
  	
  	\item \textcolor{black}{Most of \textcolor{black}{the} surveyed works are based on Open Flow protocols and their extension integrating control plane and user \textcolor{black}{plane} without \textcolor{black}{a detailed implementation \textcolor{black}{plane}}. \textcolor{black}{Moreover,} the scalability of \textcolor{black}{the} user \textcolor{black}{plane} is not considered while taking  mainly control plane into consideration~\cite{SDNFVsurvey}. Researchers have \textcolor{black}{a great} opportunity to explore regarding the standardization and scalability of these core network entities.} 
  	
  	\item \textcolor{black}{To boost spectral/energy efficiency and reduce latency utilizing CRAN and coordinated multi point (CoMP), mmWave is an attractive choice for front/back hauling because of its low implementation cost and \textcolor{black}{unavailability of} fiber everywhere~\cite{backhaulsurbey}. However, \textcolor{black}{ research in mmWave} regarding front/back hauling is \textcolor{black}{a popular} area of investigation. Dynamic, intelligent and adaptive techniques need to be developed with optimized utilization of the heterogeneous back hauling networks while catering low latency.}
  	
  	\item \textcolor{black}{Even though MEC is envisioned to reduce individual computation, inclusion of the caching in MEC can further boost users' QoE. The caching enabled MEC will provide content delivery and memory support for memory hungry applications such as VR, and online gaming along with BS level caching~\cite{wang2017survey}. \textcolor{black}{Researchers} are \textcolor{black}{encouraged} to study  various \textcolor{black}{trade-offs} such as  \textcolor{black}{ capacity versus latency, storage versus link load, and memory versus rate.} }
  	
  \end{itemize}
  
  \subsection{\textcolor{black}{Caching Issues}}
  \textcolor{black}{Edge caching can be \textcolor{black}{ a critical} tool for latency reduction along spectral and energy efficiency improvement. Recently, this issue  attracted  huge attention from \textcolor{black}{researchers} in both academia and industry resulting in many different approaches. \textcolor{black}{However}, we believe there are still extensive research problems open and need further exploration.}

  \begin{itemize}
  	\item  \textcolor{black}{Even though several exciting works have been carried out for \textcolor{black}{content} placement and content delivery time, \textcolor{black}{further studies} could be done regarding the latency aspects such as how latency is impacted by  caching size and location, and wireless channel parameters~\cite{8007071}.}
  	
  	\item \textcolor{black}{\textcolor{black}{Assuming that content delivery and content placement are the main two phases of wireless caching,} network architecture including caching storage size, placement, \textcolor{black}{and} cooperation \textcolor{black}{for} caching are potential areas for further study~\cite{coperativecaching,6736753}. Besides \textcolor{black}{that, the} protocol design for caching redundancy and intra cache communication can be investigated with latency constraint~\cite{6736753}. In this \textcolor{black}{regard}, performance limits and bounds of caching can be  studied for getting insights on optimum performance.}
  	
  	\item \textcolor{black}{\textcolor{black}{The} BS makes \textcolor{black}{a} tunnel between UE and EPC for content request. However, the contents are packeted  through GTP tunnel which creates difficulty in content-aware or \textcolor{black}{object-orientated} caching~\cite{8060515}. Proper protocol designs can address such problems.}
  	
  	\item \textcolor{black}{In low user density areas, caching capacity may be in surplus for serving UE, while  in \textcolor{black}{urban areas}, the situation may be opposite. Intelligent and co-operative resource allocation and caching strategies can ensure proper hit ratio along with low latency content delivery~\cite{6736753}. In this regard, relatively \textcolor{black}{few} works \textcolor{black}{are available} in \textcolor{black}{ the} literature, \textcolor{black}{and can be further investigated.}} 
  	
  	\item \textcolor{black}{Mobility is an \textcolor{black}{important} issue in latency critical \textcolor{black}{applications} such as AR. The movement and trajectory boost the location and performance information for local caching which handle the current user's experience. The movement of users among cells will incur interference and pilot contamination along with complication in system configurations and user-server association policies. Moreover, frequent handover will introduce latency degrading user's experience~\cite{8067654}. \textcolor{black}{Thus, handover in diverse caching scenarios with focus on low latency can be a potential area of research.} }
  	
  \end{itemize}

\section{Conclusion}

\textcolor{black}{
	Along with very large capacity, massive connection density, and ultra high reliability, 5G networks will need to support ultra low latency. The low latency will enable new services such as VR/AR, tele-medicine and tele-surgery; in some cases, latency not more than 1~ms is critical. To achieve this low latency, drastic changes in multiple network domains need to be addressed. In this paper, an extensive survey on different approaches in order to achieve low latency in 5G networks is presented. Different approaches are reviewed in the domain of RAN, core network and caching for achieving low latency. In the domain of  RAN  techniques, we have studied short frame/packets, new waveform designs, multiple access techniques, modulation and coding schemes, control channel approaches, symbol detection methods, transmission techniques, mmWave aggregation, cloud RAN, reinforcing QoS and QoE, and location aware communication as different aspects of facilitating low latency.}

	\textcolor{black}{ On the other hand, SDN, NFV and MEC/fog network architectures along with high speed backhaul are reviewed in the literature   for core network with vision to meet the low latency requirements of 5G. The new core network will provide diverse advantages such as distributed network functionality, independence of software platform from hardware platform, and separation of data plane from software plane, which will all help in latency reduction.  In caching, distributed and centralized caching with various trade-offs, cache placement and content delivery have been proposed for latency reduction in content download. Following this, promising results from field tests, trials and experiments have also been presented here. However, more practical and efficient  techniques in the presence of existing solutions need to be investigated before the standardization of 5G. \textcolor{black}{In this regard, we discussed the open issues, challenges and future research directions for researchers.} The authors believe that this survey will serve as a valuable resource for latency reduction for the emerging  5G cellular networks and beyond.}



\section*{Acknowledgment}

The work was supported by the National Science Foundation entitled “Towards Secure Networked Cyber-Physical Systems: A Theoretic Framework with Bounded Rationality” and “Cyber Physical Solution for High Penetration Renewables in Smart Grid" under the grant number 1446570 and 1553494, respectively.

\bibliographystyle{IEEEtran}
\bibliography{referkeyalimanuscript1}

\end{document}